\documentstyle[kaspp,twoside]{book}

\newcommand{\etal}{{\it et al. \,}}

\newcommand{\sunn}{$_\odot$ }

\makehead{KIM AND LEE}{NGC 185 AND NGC 205}

\begin{document}
\jkastitle{31}{1}{15}{1998}

\title{\hspace{2.6cm} SURFACE PHOTOMETRY
\newline OF THE DWARF ELLIPTICAL GALAXIES NGC 185 AND NGC 205\footnotemark[1]}

\footnotetext[1]{This paper is based in part on observations which were made 
with the Palomar 1.5m telescope which is jointly operated 
by the California Institute of Technology and 
the Carnegie Institution of Washington.}

{\center
{\large\textbf{\textsc{S}}}\textbf{\textsc{ANG}} 
{\large\textbf{\textsc{C}}}\textbf{\textsc{HUL}}
{\large\textbf{\textsc{K}}}\textbf{\textsc{IM}}
\textbf{\textsc{AND}}
{\large\textbf{\textsc{M}}}\textbf{\textsc{YUNG}}
{\large\textbf{\textsc{G}}}\textbf{\textsc{YOON}}
{\large\textbf{\textsc{L}}}\textbf{\textsc{EE}} \endcenter}
\vspace{-0.2cm}

\affil{Department of Astronomy, Seoul National University, Seoul 151-742, Korea \\
E-mail: sckim@astro.snu.ac.kr, mglee@astrog.snu.ac.kr}
\recaccp{May. 1, 1998}{May. ??, 1998}

\begin{abstract}
We present {\it BVRI} CCD surface photometry for the central($6^\prime$.35 $\times$ 
$6^\prime$.35) regions of the  dwarf elliptical galaxies NGC 185 and NGC 205 in 
the Local Group. Surface brightness profiles of NGC 185 ($R <$ 225\arcsec) and
NGC 205 ($R <$ 186\arcsec) show excess components in the central regions.
The colors of NGC 185 get bluer inward at $R <$ 25\arcsec, 
while they remain constant at $R \geq$ 25\arcsec.  
The colors of NGC 205  get bluer inward at $1\arcsec <R < 50\arcsec$, 
and remain flat outside. 
Our photometry, supplemented by the photometry based on the far-ultraviolet 
and visual images of the $HST$ archive data, shows
that there is an inversion of color at the very nucleus region 
(at about 1\arcsec). The implications of the redder color of the core part
of the nucleus compared with neighboring regions are discussed.
The amount of the excess components in the central regions of these galaxies 
is estimated to be $\approx 10^5$ $L_\odot$.
Distributions of dust clouds in the central regions of the two galaxies are also
investigated.

\end{abstract}


\section{INTRODUCTION}

NGC 185($\alpha_{2000} = 00^h 38^m 58^s.0$, $\delta_{2000} = +48^\circ 20^\prime
18\arcsec$) and NGC  205($\alpha_{2000} =  00^h 40^m  22^s.5$, $\delta_{2000}  = 
+41^\circ 41^\prime 11\arcsec$) are companion dwarf galaxies 
of M31.  They are classified as peculiar dwarf elliptical galaxies(dE3pec for NGC 185 
and S0/E5pec for NGC 205; Sandage \& Tammann 1987) due to the presence of blue stars and
dust clouds in their central regions (Lee \etal 1993, Lee 1996, 
and references therein).  

 NGC 185 and NGC 205 are easily resolved even from the ground observations so that it is possible
to study individual stars in these galaxies (for example, see Baade 1944, 1951; Lee \etal 1993; and Lee 1996). 
However, severe crowding prevents us from studying faint stars in these galaxies in detail.
We have to rely on surface photometry to understand the populations of faint stars.
There have been several surface photometry studies for these two 
galaxies(Hodge 1963, Price 1985,  Kent 1987, and Lee  \etal 1993 for NGC  185; and 
Hodge 1973, Price \& Grasdalen 1983, Kent 1987, Peletier 1993, Jones \etal 1996, 
and Lee 1996 for NGC 205).  
Old studies among these are based on photographic and/or photoelectric photometry
so that they suffer from large errors for the outer regions of these galaxies.
Recent studies among these are based on CCD observations(Kent 1987 
and Lee \etal 1993 for NGC 185, and Peletier 1993, Lee 1996 and Jones \etal 1996
for NGC 205).
Kent (1987) presented Gunn {\it r} band surface photometry of NGC 185 and NGC 205.
Peletier (1993) presented, for the first time, multi-band ({\it UBRI}) CCD photometry of NGC 205, 
but {\it V}-band was not included in this study.
Lee \etal (1993) and Lee (1996) presented {\it BVRI} photometry of NGC 185 and NGC 205, but these
studies were limited to a small field.
Jones \etal (1996) presented {\it V}-band photometry of NGC 205 based on $HST$ observations, 
but it is limited to the nucleus region at $R \leq 1\arcsec.15$.
Thus these recent studies are limited either to one-band photometry or to a small field.

Especially Lee \etal (1993) and Lee (1996) found that the colors get redder outward at $R>50\arcsec$
for both NGC 185 and NGC 205, while Peletier (1993) showd that the colors of NGC 205 get redder
outward at $R<50''$, but remain almost constant at $R>50\arcsec$.
It remains to be investigated whether the colors of these galaxies will 
get continously redder outward or not.

In this paper, we present surface photometry based on {\it BVRI} CCD images 
for the central $6\arcmin.35 \times 6\arcmin.35$ areas of NGC 185 and NGC 205.
We have derived surface brightness distributions, color distributions, and structural parameters 
for $R <$ 225$\arcsec$ of NGC 185 and $R <$ 186$\arcsec$ of NGC 205. 
We also have investigated the color profiles of the nucleus of NGC 205 using the 
far-ultraviolet and visual images in the $HST$ archive data. 
A brief progress report of this study was given in Kim \& Lee (1996).
This paper is composed as follows:  Section II describes observations and 
data reduction and Sec. III presents surface photometry results.  
Section IV discusses light excesses in the central regions of NGC 185 and NGC 205,
color gradient in the nuclear region of NGC 205,
and dust clouds in the two galaxies.
Finally, the primary results are summarized in Sec. V.

\section{OBSERVATIONS AND DATA REDUCTION}
\subsection{Observations}

  {\it BVRI} CCD images of NGC 185 were obtained at the Palomar 1.5m telescope 
using the Tektronics  1024 $\times$ 1024  pixels CCD camera on December 8, 1991(UT) and  
{\it VRI} CCD images of NGC 205  were obtained using the same equipment on August  8, 1992(UT).  
The  pixel scale  is 0\arcsec.376 per pixel and the size of the CCD field is 
$6\arcmin.35$ $\times$ $6\arcmin.35$.  
The observing log is given in Table 1. 

\vspace{0.3cm}
\begin{center}
\centerline{{\bf Table 1.} Journal of observations for NGC 185 and NGC 205.} \smallskip
\begin{tabular}{cccccc}
\tableline
\tableline
Object & Filter & Exposure Time & UT(start) & Seeing & Airmass \\
\tableline
NGC 185 &$B$& 1800 sec& 8 Dec 91 $5 ^{h} 58^{m}$&  2\arcsec.2& 1.19\\
        &$V$&  900 sec& 8 Dec 91 $4 ^{h} 43^{m}$&  2\arcsec.7& 1.07\\
        &$R$&  900 sec& 8 Dec 91 $5 ^{h} 26^{m}$&  2\arcsec.3& 1.13\\
        &$I$&  600 sec& 8 Dec 91 $5 ^{h} 09^{m}$&  1\arcsec.8& 1.10\\
\tableline
NGC 205 &$V$&  900 sec& 8 Aug 92 $12 ^{h} 15^{m}$& 1\arcsec.7& 1.03\\
        &$R$&  600 sec& 8 Aug 92 $11 ^{h} 43^{m}$& 1\arcsec.3& 1.01\\
        &$I$&  600 sec& 8 Aug 92 $11 ^{h} 55^{m}$& 1\arcsec.3& 1.02\\
\tableline
\end{tabular}
\end{center}
\vspace{0.3cm}

We have also reanalyzed the CFHT {\it BVRI} CCD images of NGC 185 and NGC 205 
used in Lee \etal (1993) and Lee (1996), respectively, to supplement our data.
We have created final data by combining the Palomar data for $R>20\arcsec$ and the CFHT data
for $R<20\arcsec$. 
  Greyscale maps of {\it V} CCD images of NGC 185 and NGC 205 obtained at the 
Palomar 1.5m telescope are shown in Fig. 1.

\noindent
\subsection{Data Reduction}

We have derived surface photometry of the galaxies from the CCD images 
using an ellipse fitting task ELLIPSE in IRAF
\footnote[1]{IRAF is distributed by the National
Optical Astronomy Observatories, which is operated by the Association of
Universities for Research in Astronomy, Inc., under cooperative
agreement with the National Science Foundation.}/STSDAS.
We have also used two other methods for surface photometry 
to check our results: 
SPIRAL (ellipse fitting package developed at Kiso Observatory(Ichikawa \etal 1987)) 
installed in IRAF and
elliptical annulus aperture photometry using POLYPHOT in IRAF. 
The results from the three methods agree very well. We have used
the results based on ELLIPSE for final analysis.

\vspace{0.5cm}
\begin{minipage}[t]{8.0cm}
   \epsfxsize=8.0cm
   \epsfysize=8.0cm
   \epsfbox{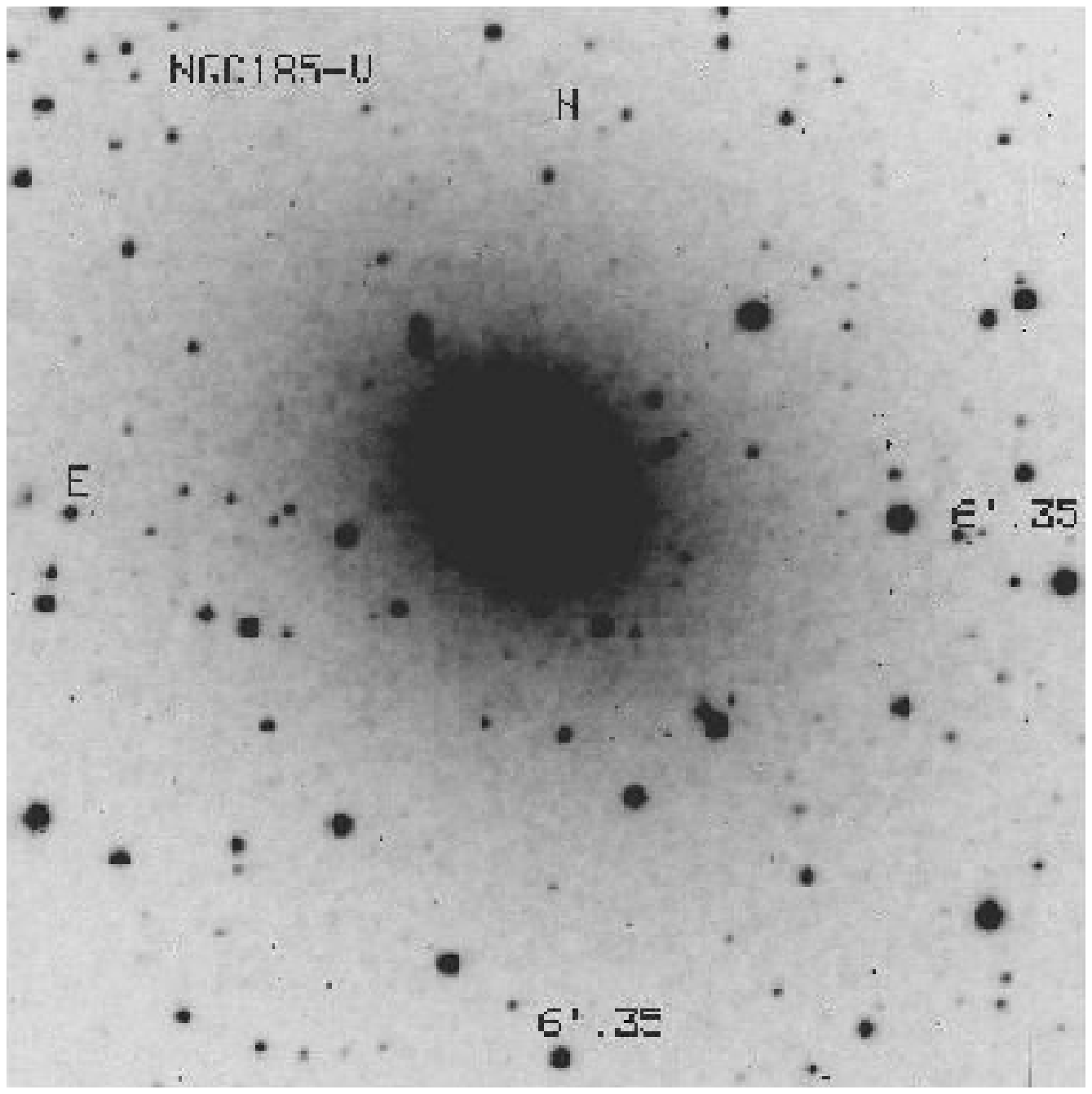}
\end{minipage} \hfill
\begin{minipage}[t]{8.0cm}
   \epsfxsize=8.0cm
   \epsfysize=8.0cm
   \epsfbox{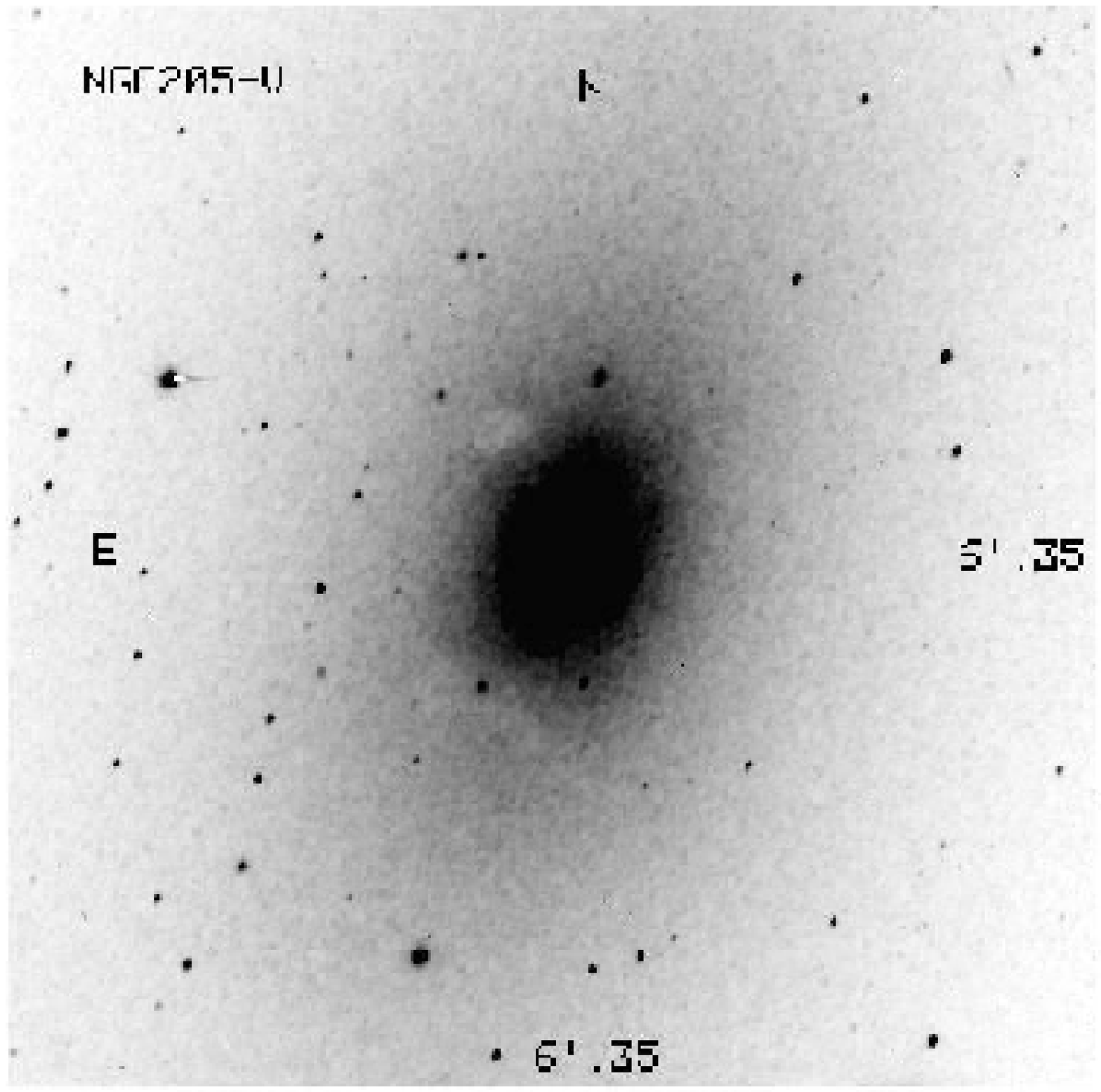}
\end{minipage}

(a) \hspace{8.3cm} (b)

\hspace{-0.8cm} {\footnotesize \parbox[t]{17.5cm}{{\bf Fig. 1.}  
A {\it V}-band CCD image of NGC 185 {\bf (a)} and NGC 205 {\bf (b)}.  
North is at the top and east is to the
 left.  The scale is 0\arcsec.376 per pixel and the size of the
 field is $6 \arcmin .35 \times 6 \arcmin .35$.}}
\vspace{1cm}

The instrumental magnitudes were transformed onto the standard system using
the photometry given by Lee \etal (1993) for NGC 185 and using the photometry
of standard stars (Landolt 1983) observed on the same night for NGC 205.
The errors for the standard transformation are $\sim$0.03 mag.

  We have adopted as the distances to NGC 185 and NGC 205,
$620 \pm 60$ kpc ($(m-M)_0 = 23.96 \pm 0.21$ mag, Lee \etal 1993) and 
$830 \pm 115$ kpc ($(m-M)_0=24.6 \pm 0.3$ mag, Lee 1996), respectively, in this study.  
At these distances, 1$\arcsec$ corresponds 
to 3.01 pc and 4.02 pc for NGC 185 and NGC 205, respectively.

\section{RESULTS}

\subsection{Surface Brightness Profiles}
\centerline{\bf 1) NGC 185}

Fig. 2(a) and 2(b) show surface brightness profiles for the region of 
$R <$ 225$\arcsec$ of NGC 185, which are also listed in Table 2.
The radius is given in terms of effective radius along 
the semi-major axis from the center of the galaxy, and
the surface brightness is given in units of mag per square arcsecond.
Fig. 2 shows that surface brightness profiles of NGC 185 are flat 
in the central region and decrease smoothly outward.
We plotted the photometric errors only for $B$-band to show the size of typical errors.

We have also plotted the results from the previous studies 
for comparison in Fig. 2(Hodge 1963, Kent 1987, and Lee \etal 1993).   
Hodge (1963)'s {\it V} surface photometry profile 
agrees well with that of this study 
for 10\arcsec $< R <$ 50$\arcsec$ 
but show significant difference for $R < 10 \arcsec$ and $R > 50$\arcsec.  
Gunn $r$ photometry of Kent (1987) is converted to Cousins 
$R$ photometry using the approximate relation $R=r-0.31$ (Barsony 1989, Lee \etal 1993).  
His photometry shows reasonably good agreement with ours for $R < 135 \arcsec$.  
{\it BVRI} photometry of Lee \etal (1993) shows overall  agreement with ours except 
for the brightening bump at $R \sim 12 \arcsec$ in all bands.  
It appears that the bump in Lee \etal (1993) is due to a bright foreground star of {\it V}=16.03 mag 
and $B-V$=0.77 mag (star \# 2441  of Lee  \etal 1993), which was not removed 
in the process of photometry in Lee \etal (1993).  

\begin{minipage}[t]{13.6cm}
   \epsfxsize=8.7cm
   \epsfysize=13.6cm
   \epsfbox{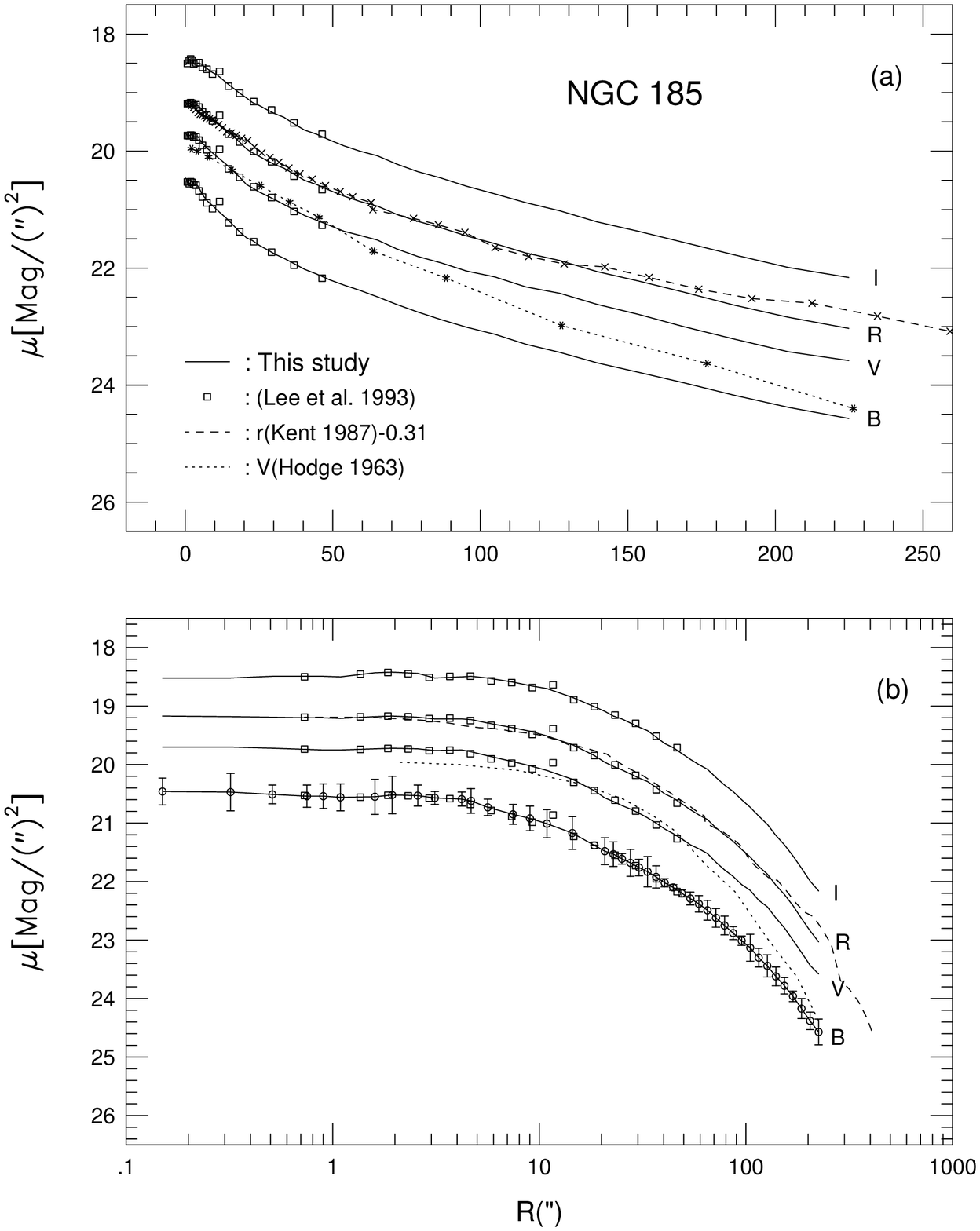}
\end{minipage} \hspace{-0.01cm}
\begin{minipage}[t]{13.6cm}
   \epsfxsize=8.7cm
   \epsfysize=13.6cm
   \epsfbox{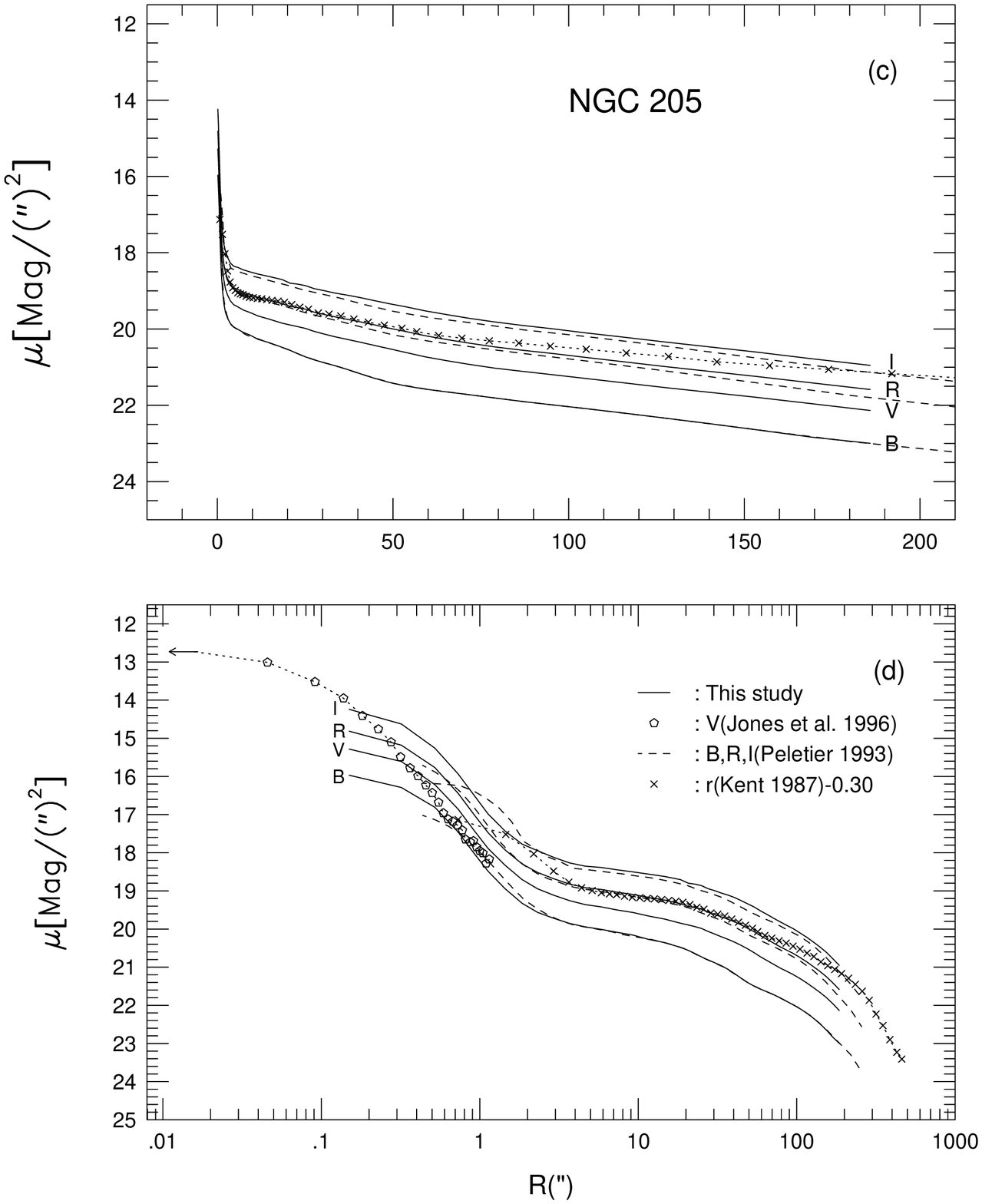}
\end{minipage}
\hspace{-0.8cm} {\footnotesize \parbox[t]{17.5cm}{{\bf Fig. 2. (a)} 
Surface brightness profiles of NGC 185 vs effective radius
(in arcsec) obtained in this study (solid lines).
The {\it V} photometry results of Hodge (1963) are shown by a dotted line.
Gunn $r$ photometry of Kent(1987) is transformed to Cousins $R$
magnitude and is represented by a dashed line.
{\it BVRI} CCD data of Lee \etal (1993) are plotted by open squares.
{\bf (b)} Same as in (a) but the abscissa is in logarithm
of the effective radius.  
The $B$ surface brightness errors are plotted.
{\bf (c)} Surface brightness profiles of NGC 205 vs effective
radius (in arcsec) obtained in this study (solid lines).
Gunn $r$ photometry of Kent (1987) is transformed to 
Cousins $R$ magnitude and is represented by $\times$ symbol and a dotted line.
$BRI$ CCD photometry data of Peletier (1993) are shown by dashed lines.
{\bf (d)} Same as in (c) but the abscissa is given in terms of logarithm
of the effective radius.  
{\it V} photometry of Jones \etal (1996) based on the $HST$ data are shown 
by pentagons in the region of $R \leq$ 1\arcsec.15, and the arrow
symbol indicates the central value.}}
\vspace{1cm}

\centerline{\bf 2) NGC 205}
Fig. 2(c) and 2(d) show surface brightness profiles 
for $R <$ 186$\arcsec$ of NGC 205, which are also listed in Table 2.  
Because the Palomar data lacks the $B$ band data, 
we used  the $B$ band  results of Peletier (1993) for $R > 20 \arcsec$.  
Fig. 2(c) and 2(d) show that the surface brightness profiles of NGC 205 
have a rapidly brightening part in the nuclear region, 
a flat part in the central region and
a smoothly decreasing part with radius in the outer region.

  In Fig. 2(c) and (d), we have also showed for comparison the previous 
surface photometry results of Kent (1987), Peletier (1993), and Jones \etal (1996).  
Gunn  $r$ photometry of Kent (1987) is converted to Cousins $R$  photometry 
using  $R=r-0.30$ (Barsony 1989, Lee 1996).  
Even though his data show good agreement with ours for 4\arcsec $<$ $R < $ 60$\arcsec$, 
they get much brighter than ours at $R > $ 60$\arcsec$ and 
diverge significantly from ours at $R < 4$\arcsec.  
Peletier (1993)'s $BRI$ data also show, in general, good agreement with ours 
for $R > 3 \arcsec$ except that his $R$ and $I$ mags are 
systematically $\sim$0.1 mag fainter than ours.  
The {\it V} surface brightness profiles 
of the nuclear region($R \leq 1$\arcsec.15) of NGC 205 
obtained by Jones \etal (1996) using $HST$ are also shown in Fig. 2(d).  
Their {\it V} surface brightness profiles are less bright than ours for $R \sim 1 \arcsec$ 
but get significantly brighter than ours at $R <$ 0\arcsec.3, 
giving $\mu_{V}(0) = 12.73$ mag arcsec$^{-2}$.

{\scriptsize
\begin{center}
{\normalsize \centerline{{\bf Table 2.} Surface brightness profiles of NGC 185 and NGC 205.}}
\begin{tabular}{rcccccccc|rccccccc}
\tableline
\tableline
NGC 185 &&&&&&&&&    NGC 205 &&&&&&&\\
\tableline
$R$(\arcsec)&  $\mu_B$&  $\sigma_B$&  $\mu_V$&  $\sigma_V$&  
$\mu_R$&  $\sigma_R$&  $\mu_I$&  $\sigma_I$&
$R$(\arcsec)&  $\mu_B$&  $\mu_V$&  $\sigma_V$&
$\mu_R$&  $\sigma_R$&  $\mu_I$&  $\sigma_I$ \\
\tableline
 0.15& 20.46& 0.23& 19.70& 0.13& 19.17& 0.12& 18.52& 0.02&  0.15& 15.97& 15.28& 0.00& 14.81& 0.00& 14.24& 0.00\\ 
 0.32& 20.47& 0.32& 19.70& 0.21& 19.18& 0.16& 18.52& 0.05&  0.32& 16.29& 15.61& 0.00& 15.18& 0.00& 14.62& 0.00\\ 
 0.51& 20.51& 0.16& 19.72& 0.17& 19.19& 0.10& 18.49& 0.08&  0.52& 16.81& 16.21& 0.00& 15.77& 0.00& 15.26& 0.00\\ 
 0.75& 20.54& 0.19& 19.74& 0.28& 19.20& 0.18& 18.49& 0.10&  0.73& 17.46& 16.86& 0.00& 16.44& 0.00& 15.93& 0.00\\ 
 0.90& 20.54& 0.21& 19.75& 0.23& 19.20& 0.16& 18.49& 0.12&  0.93& 18.04& 17.44& 0.00& 17.02& 0.00& 16.52& 0.00\\ 
 1.09& 20.56& 0.23& 19.75& 0.24& 19.21& 0.14& 18.50& 0.13&  1.13& 18.49& 17.89& 0.00& 17.48& 0.00& 16.99& 0.00\\ 
 1.60& 20.55& 0.30& 19.73& 0.27& 19.18& 0.19& 18.43& 0.17&  1.44& 18.94& 18.32& 0.00& 17.92& 0.00& 17.43& 0.00\\ 
 1.94& 20.52& 0.32& 19.72& 0.27& 19.17& 0.19& 18.42& 0.17&  1.85& 19.33& 18.70& 0.00& 18.29& 0.00& 17.77& 0.00\\ 
 2.58& 20.53& 0.18& 19.73& 0.17& 19.19& 0.17& 18.44& 0.17&  2.32& 19.57& 18.95& 0.00& 18.52& 0.00& 17.95& 0.00\\ 
 3.12& 20.57& 0.11& 19.76& 0.13& 19.22& 0.17& 18.52& 0.12&  2.93& 19.72& 19.11& 0.00& 18.68& 0.00& 18.12& 0.00\\ 
 4.21& 20.59& 0.12& 19.75& 0.16& 19.22& 0.18& 18.50& 0.13&  3.69& 19.85& 19.26& 0.00& 18.82& 0.00& 18.27& 0.00\\ 
 4.67& 20.62& 0.21& 19.78& 0.18& 19.23& 0.20& 18.49& 0.14&  4.64& 19.94& 19.36& 0.00& 18.92& 0.00& 18.35& 0.00\\ 
 5.63& 20.73& 0.14& 19.85& 0.19& 19.29& 0.20& 18.52& 0.15&  5.84& 20.01& 19.42& 0.00& 18.96& 0.00& 18.39& 0.00\\ 
 7.46& 20.85& 0.17& 19.95& 0.23& 19.37& 0.24& 18.59& 0.14&  7.35& 20.09& 19.50& 0.00& 19.03& 0.00& 18.45& 0.00\\ 
 9.01& 20.92& 0.21& 20.02& 1.99& 19.43& 1.87& 18.65& 0.14&  9.26& 20.16& 19.57& 0.00& 19.09& 0.00& 18.50& 0.00\\ 
10.88& 21.01& 0.24& 20.10& 0.30& 19.51& 0.27& 18.70& 0.19& 11.65& 20.27& 19.66& 0.00& 19.17& 0.00& 18.56& 0.00\\ 
14.44& 21.17& 0.28& 20.25& 0.37& 19.66& 0.32& 18.85& 0.21& 14.67& 20.35& 19.74& 0.00& 19.24& 0.00& 18.62& 0.00\\ 
18.47& 21.38& 0.00& 20.45& 0.00& 19.84& 0.00& 19.01& 0.00& 18.47& 20.48& 19.85& 0.00& 19.34& 0.00& 18.70& 0.00\\ 
20.75& 21.48& 0.23& 20.57& 0.12& 19.96& 0.13& 19.09& 0.08& 20.75& 20.56& 19.90& 0.10& 19.41& 0.11& 18.80& 0.10\\ 
22.83& 21.53& 0.21& 20.63& 0.12& 20.03& 0.13& 19.17& 0.07& 22.83& 20.64& 19.94& 0.13& 19.44& 0.14& 18.82& 0.09\\ 
25.11& 21.61& 0.09& 20.70& 0.07& 20.09& 0.09& 19.22& 0.09& 25.11& 20.73& 19.99& 0.14& 19.48& 0.15& 18.86& 0.10\\ 
27.62& 21.68& 0.23& 20.76& 0.13& 20.16& 0.15& 19.29& 0.11& 27.62& 20.81& 20.06& 0.16& 19.56& 0.17& 18.94& 0.09\\ 
30.38& 21.76& 0.14& 20.83& 0.10& 20.23& 0.13& 19.36& 0.12& 30.38& 20.88& 20.13& 0.15& 19.62& 0.17& 18.99& 0.09\\ 
33.42& 21.83& 0.26& 20.91& 0.18& 20.29& 0.21& 19.42& 0.20& 33.42& 20.96& 20.20& 0.15& 19.68& 0.19& 19.06& 0.08\\ 
36.76& 21.92& 0.19& 21.00& 0.15& 20.39& 0.20& 19.53& 0.09& 36.76& 21.05& 20.27& 0.16& 19.74& 0.23& 19.11& 0.08\\ 
40.44& 22.02& 0.07& 21.10& 0.05& 20.50& 0.06& 19.64& 0.08& 40.44& 21.17& 20.34& 0.21& 19.81& 0.22& 19.17& 0.07\\ 
44.48& 22.10& 0.05& 21.18& 0.04& 20.57& 0.06& 19.71& 0.08& 44.48& 21.28& 20.42& 0.06& 19.89& 0.06& 19.25& 0.07\\ 
48.93& 22.20& 0.06& 21.27& 0.04& 20.67& 0.06& 19.81& 0.07& 48.93& 21.40& 20.52& 0.09& 19.98& 0.07& 19.34& 0.07\\ 
53.82& 22.29& 0.11& 21.37& 0.07& 20.77& 0.08& 19.91& 0.09& 53.82& 21.50& 20.62& 0.07& 20.07& 0.06& 19.42& 0.07\\ 
59.20& 22.38& 0.14& 21.44& 0.15& 20.86& 0.10& 20.00& 0.10& 59.20& 21.58& 20.73& 0.09& 20.18& 0.07& 19.52& 0.08\\ 
65.12& 22.49& 0.17& 21.52& 0.25& 20.94& 0.20& 20.08& 0.21& 65.12& 21.65& 20.83& 0.11& 20.27& 0.09& 19.62& 0.09\\ 
71.63& 22.62& 0.16& 21.67& 0.15& 21.08& 0.14& 20.22& 0.13& 71.63& 21.72& 20.93& 0.08& 20.37& 0.07& 19.72& 0.08\\ 
78.80& 22.75& 0.16& 21.80& 0.13& 21.20& 0.11& 20.35& 0.11& 78.80& 21.81& 21.02& 0.06& 20.47& 0.07& 19.81& 0.08\\ 
86.68& 22.88& 0.11& 21.91& 0.15& 21.31& 0.14& 20.46& 0.14& 86.68& 21.90& 21.11& 0.07& 20.56& 0.07& 19.90& 0.07\\ 
95.35& 23.01& 0.08& 22.05& 0.08& 21.45& 0.08& 20.60& 0.09& 95.35& 21.99& 21.20& 0.07& 20.64& 0.08& 19.99& 0.08\\ 
104.88& 23.13& 0.23& 22.15& 0.29& 21.58& 0.16& 20.73& 0.15& 104.88& 22.09& 21.30& 0.08& 20.75& 0.08& 20.10& 0.09\\ 
115.37& 23.30& 0.16& 22.32& 0.21& 21.74& 0.10& 20.88& 0.11& 115.37& 22.20& 21.41& 0.09& 20.86& 0.09& 20.21& 0.09\\ 
126.91& 23.44& 0.19& 22.43& 0.31& 21.87& 0.14& 21.02& 0.16& 126.91& 22.33& 21.53& 0.09& 20.98& 0.09& 20.33& 0.10\\ 
139.60& 23.62& 0.16& 22.62& 0.23& 22.06& 0.12& 21.21& 0.13& 139.60& 22.48& 21.66& 0.11& 21.11& 0.10& 20.46& 0.11\\ 
153.56& 23.78& 0.14& 22.79& 0.21& 22.22& 0.12& 21.37& 0.13& 153.56& 22.64& 21.79& 0.14& 21.25& 0.13& 20.61& 0.14\\ 
168.92& 23.96& 0.09& 23.00& 0.09& 22.41& 0.07& 21.56& 0.09& 168.92& 22.83& 21.95& 0.14& 21.41& 0.14& 20.78& 0.15\\ 
185.81& 24.17& 0.17& 23.21& 0.23& 22.63& 0.13& 21.77& 0.15& 185.81& 23.00& 22.14& 0.16& 21.59& 0.16& 20.95& 0.17\\ 
204.39& 24.38& 0.15& 23.43& 0.14& 22.84& 0.10& 21.99& 0.13&    &      &      &      &      &      &      &\\ 
224.83& 24.57& 0.22& 23.58& 0.26& 23.03& 0.18& 22.16& 0.21&    &      &      &      &      &      &      &\\
\tableline
\end{tabular}
\end{center}
}

\subsection{Color Profiles}
\centerline{\bf 1) NGC 185}
  The color profiles of NGC 185 are shown in Fig. 3(a) and listed in Table 3.  
Fig. 3(a) shows that all colors get bluer inward at $R <$ 25$\arcsec$ and 
remain nearly flat at $R >$ 25$\arcsec$. 
In Fig. 3(a), we have also plotted for comparison 
the results of the previous studies by Hodge (1963) and Lee \etal (1993).  
The  $B-V$ data  of Hodge (1963) are on average $\sim$0.2  mag redder than ours,
while the data of Lee \etal (1993) show overall agreement with ours.  
All three colors containing $I$ band in Lee \etal (1993)'s data 
show radial gradients up to $50\arcsec$, while ours show little gradient
beyond $R>20\arcsec$.
The gradients shown for the outer region in Lee \etal appear
 to have been caused due to their incorrect sky subraction for the $I$
image.

\hspace{2cm}
\begin{minipage}[b]{13cm}
   \epsfxsize=13cm
   \epsfysize=22cm
   \epsfbox{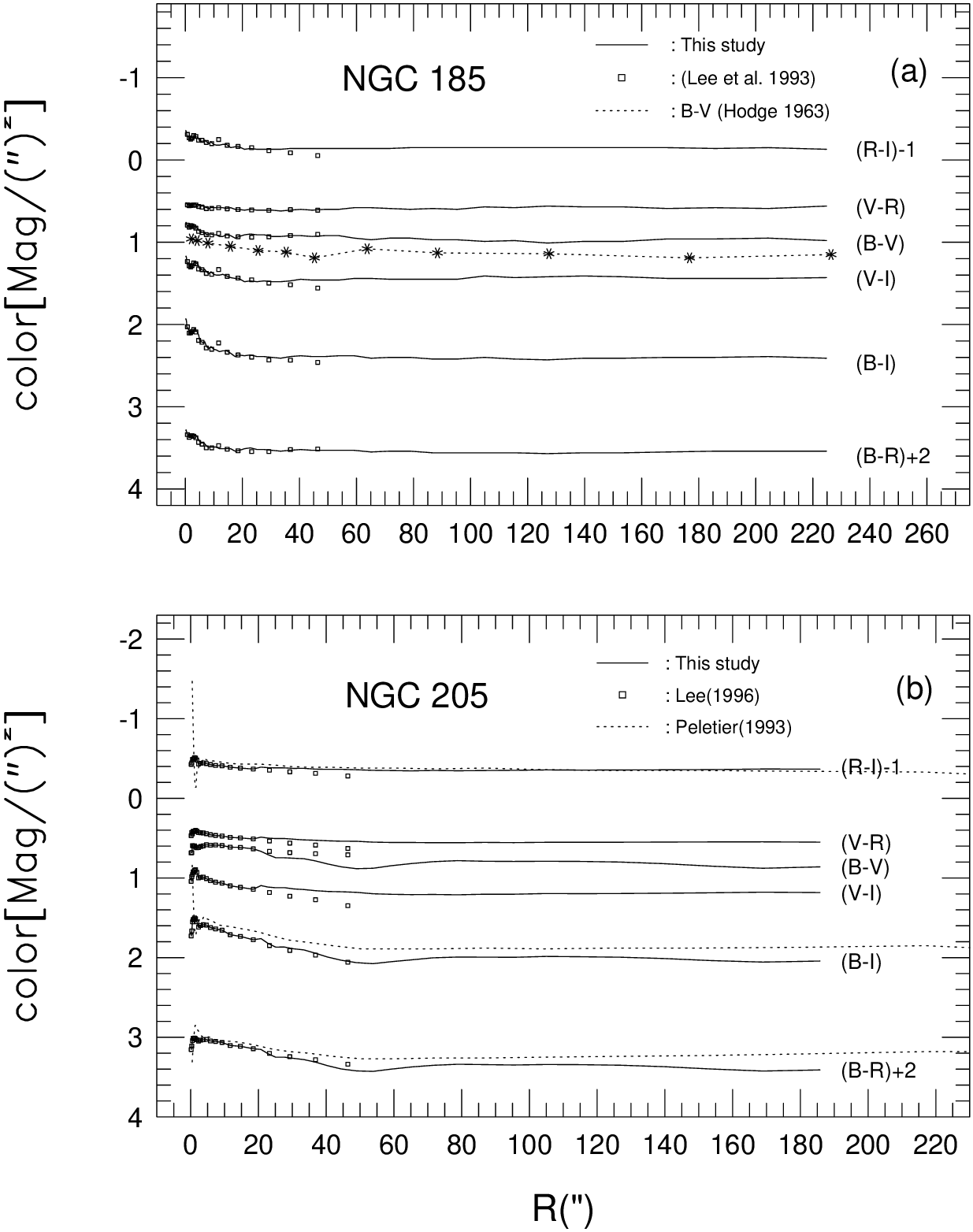}
\end{minipage} \hfill

\hspace{-0.8cm} {\footnotesize \parbox[t]{17.5cm}{{\bf Fig. 3. (a)} 
Differential color profiles vs effective radius of NGC 185 
are shown by solid lines.
The  $B-V$ color of Hodge (1963) is represented by a dotted line, and
the colors from {\it BVRI} photometry of Lee \etal (1993)
are shown by open squares.
{\bf (b)} Differential color profiles vs effective radius of NGC 205 
are shown by solid lines.
The colors of Peletier (1993) are shown by dotted lines and 
those of Lee (1996) by open squares.}}
\vspace{0.5cm}

\centerline{\bf 2) NGC 205}
 Color profiles of NGC 205 are shown in Fig. 3(b) and listed in Table 3.  
Fig. 3(b) shows:
(a) that the colors of the nucleus ($R<1\arcsec.1$) are bluer than those of the outside
and that they get redder inward. 
This color inversion will be discussed in detail in Section IV;
(b) that all the colors get bluer inward  in the central region  
($1\arcsec.1 < R < 50$\arcsec);
and (c) that they remain nearly flat in the outer region ($R > 50$\arcsec).
We have compared our results with those of Peletier (1993) and Lee (1996) in Fig. 3(b).  
A comparison of the $R-I$ color profiles of our study and Peletier (1993)'s 
shows excellent agreement except in the nuclear region.  
The $B-I$ and $B-R$ colors of Peletier's are systemically
$\approx$0.2 mag bluer than ours.  

{\scriptsize
\begin{center}
{\normalsize \centerline
   {{\bf Table 3.} Differential color profiles of NGC 185 and NGC 205.}}
\begin{tabular}{rcccccc|rcccccc}
\tableline
\tableline
NGC 185 &&&&&&& NGC 205 &&&&&&\\
\tableline
$R$(\arcsec)& $B-V$& $\sigma_{B-V}$& 
$V-R$& $\sigma_{V-R}$& $V-I$& $\sigma_{V-I}$&
$R$(\arcsec)& $B-V$& $\sigma_{B-V}$& 
$V-R$& $\sigma_{V-R}$& $V-I$& $\sigma_{V-I}$\\
\tableline
  0.15& 0.76& 0.26& 0.53& 0.18& 1.17& 0.13&   0.15& 0.69& 0.00& 0.47& 0.00& 1.04& 0.00\\ 
  0.32& 0.76& 0.38& 0.52& 0.27& 1.18& 0.22&   0.32& 0.68& 0.00& 0.43& 0.00& 0.98& 0.00\\ 
  0.51& 0.79& 0.23& 0.53& 0.20& 1.22& 0.19&   0.52& 0.59& 0.00& 0.45& 0.00& 0.96& 0.00\\ 
  0.75& 0.79& 0.34& 0.54& 0.33& 1.25& 0.30&   0.73& 0.60& 0.00& 0.41& 0.00& 0.92& 0.00\\ 
  0.90& 0.79& 0.32& 0.54& 0.28& 1.26& 0.26&   0.93& 0.60& 0.00& 0.41& 0.00& 0.92& 0.00\\ 
  1.09& 0.80& 0.33& 0.55& 0.28& 1.25& 0.28&   1.13& 0.60& 0.00& 0.41& 0.00& 0.90& 0.00\\ 
  1.60& 0.82& 0.40& 0.55& 0.33& 1.31& 0.31&   1.44& 0.62& 0.00& 0.40& 0.00& 0.90& 0.00\\ 
  1.94& 0.80& 0.42& 0.55& 0.33& 1.30& 0.32&   1.85& 0.62& 0.00& 0.41& 0.00& 0.93& 0.00\\ 
  2.58& 0.80& 0.25& 0.54& 0.24& 1.29& 0.24&   2.32& 0.62& 0.00& 0.43& 0.00& 1.00& 0.00\\ 
  3.12& 0.81& 0.17& 0.54& 0.21& 1.24& 0.17&   2.93& 0.61& 0.00& 0.43& 0.00& 0.99& 0.00\\ 
  4.21& 0.83& 0.20& 0.54& 0.24& 1.25& 0.21&   3.69& 0.60& 0.00& 0.43& 0.00& 0.99& 0.00\\ 
  4.67& 0.85& 0.28& 0.55& 0.27& 1.29& 0.23&   4.64& 0.58& 0.00& 0.44& 0.00& 1.01& 0.00\\ 
  5.63& 0.88& 0.23& 0.56& 0.28& 1.33& 0.24&   5.84& 0.59& 0.00& 0.46& 0.00& 1.03& 0.00\\ 
  7.46& 0.91& 0.28& 0.58& 0.33& 1.36& 0.27&   7.35& 0.59& 0.00& 0.47& 0.00& 1.05& 0.00\\ 
  9.01& 0.90& 2.00& 0.58& 2.73& 1.36& 2.00&   9.26& 0.59& 0.00& 0.47& 0.00& 1.06& 0.00\\ 
 10.88& 0.91& 0.39& 0.58& 0.40& 1.40& 0.35&  11.65& 0.61& 0.00& 0.49& 0.00& 1.10& 0.00\\ 
 14.44& 0.91& 0.46& 0.59& 0.49& 1.41& 0.43&  14.67& 0.62& 0.00& 0.50& 0.00& 1.12& 0.00\\ 
 18.47& 0.93& 0.00& 0.60& 0.00& 1.44& 0.00&  18.47& 0.63& 0.00& 0.51& 0.00& 1.14& 0.00\\ 
 20.75& 0.91& 0.26& 0.61& 0.17& 1.48& 0.14&  20.75& 0.67& 0.10& 0.49& 0.15& 1.10& 0.14\\ 
 22.83& 0.90& 0.24& 0.60& 0.17& 1.47& 0.14&  22.83& 0.71& 0.13& 0.50& 0.19& 1.11& 0.16\\ 
 25.11& 0.91& 0.11& 0.61& 0.11& 1.48& 0.12&  25.11& 0.75& 0.14& 0.50& 0.20& 1.12& 0.17\\ 
 27.62& 0.91& 0.27& 0.61& 0.20& 1.47& 0.17&  27.62& 0.75& 0.16& 0.50& 0.23& 1.12& 0.18\\ 
 30.38& 0.93& 0.17& 0.61& 0.16& 1.48& 0.15&  30.38& 0.75& 0.15& 0.51& 0.23& 1.14& 0.18\\ 
 33.42& 0.93& 0.31& 0.62& 0.28& 1.48& 0.27&  33.42& 0.76& 0.15& 0.52& 0.25& 1.14& 0.17\\ 
 36.76& 0.92& 0.24& 0.61& 0.25& 1.47& 0.17&  36.76& 0.79& 0.16& 0.53& 0.28& 1.16& 0.18\\ 
 40.44& 0.92& 0.08& 0.60& 0.08& 1.45& 0.09&  40.44& 0.82& 0.21& 0.53& 0.31& 1.17& 0.23\\ 
 44.48& 0.93& 0.07& 0.61& 0.08& 1.46& 0.09&  44.48& 0.86& 0.06& 0.54& 0.09& 1.17& 0.10\\ 
 48.93& 0.92& 0.07& 0.60& 0.07& 1.46& 0.08&  48.93& 0.88& 0.09& 0.54& 0.11& 1.18& 0.11\\ 
 53.82& 0.92& 0.13& 0.60& 0.11& 1.46& 0.11&  53.82& 0.88& 0.07& 0.55& 0.09& 1.20& 0.09\\ 
 59.20& 0.95& 0.21& 0.58& 0.18& 1.44& 0.18&  59.20& 0.85& 0.09& 0.55& 0.11& 1.21& 0.12\\ 
 65.12& 0.97& 0.30& 0.58& 0.32& 1.44& 0.33&  65.12& 0.82& 0.11& 0.55& 0.14& 1.21& 0.14\\ 
 71.63& 0.95& 0.22& 0.59& 0.20& 1.45& 0.19&  71.63& 0.80& 0.08& 0.56& 0.11& 1.21& 0.11\\ 
 78.80& 0.95& 0.20& 0.60& 0.17& 1.45& 0.17&  78.80& 0.78& 0.06& 0.56& 0.09& 1.21& 0.10\\ 
 86.68& 0.97& 0.19& 0.59& 0.20& 1.45& 0.21&  86.68& 0.79& 0.07& 0.55& 0.10& 1.20& 0.10\\ 
 95.35& 0.97& 0.11& 0.60& 0.11& 1.45& 0.12&  95.35& 0.79& 0.07& 0.56& 0.10& 1.20& 0.11\\ 
104.88& 0.99& 0.38& 0.57& 0.34& 1.41& 0.33& 104.88& 0.79& 0.08& 0.55& 0.11& 1.20& 0.12\\ 
115.37& 0.98& 0.26& 0.58& 0.23& 1.43& 0.23& 115.37& 0.79& 0.09& 0.55& 0.12& 1.20& 0.13\\ 
126.91& 1.01& 0.36& 0.56& 0.34& 1.42& 0.35& 126.91& 0.80& 0.09& 0.55& 0.13& 1.20& 0.14\\ 
139.60& 0.99& 0.28& 0.57& 0.26& 1.41& 0.26& 139.60& 0.82& 0.11& 0.55& 0.15& 1.19& 0.16\\ 
153.56& 0.99& 0.25& 0.57& 0.24& 1.42& 0.24& 153.56& 0.85& 0.14& 0.55& 0.19& 1.19& 0.20\\ 
168.92& 0.96& 0.13& 0.59& 0.11& 1.44& 0.12& 168.92& 0.88& 0.14& 0.55& 0.20& 1.18& 0.20\\ 
185.81& 0.96& 0.29& 0.58& 0.27& 1.44& 0.28& 185.81& 0.86& 0.16& 0.55& 0.23& 1.18& 0.24\\ 
204.39& 0.95& 0.20& 0.59& 0.17& 1.44& 0.19&&&&&&\\ 
224.83& 0.98& 0.34& 0.56& 0.31& 1.43& 0.33&&&&&&\\ 
\tableline
\end{tabular}
\end{center}
}

\subsection{Structural Parameters}
\centerline{\bf 1) NGC 185}

Ellipticity($\epsilon = 1 - b/a$) and position angle(PA; N through E) 
distributions of NGC 185 are similar in {\it B, V, R,} and {\it I} images.
For simplicity we have plotted only the $R$ band  profiles of $\epsilon$ and PA 
in Fig. 4(a) and (b), which are listed in Table 4.  
Fig. 4(a) and 4(b) show that
in the region of $R >25$\arcsec, $\epsilon$ is $\sim$0.2 and PA is $\sim 50^\circ$, 
while in $R <25$\arcsec, $\epsilon$  and PA vary significantly
($0.04 < \epsilon < 0.42$ and $8^\circ <PA< 100^\circ$).  
The inner region ($R <$ 25\arcsec) in which $\epsilon$ and PA vary much is 
where young stars and eminent dust clouds are concentrated. 
The $\epsilon$ distribution of Hodge (1963) in Fig. 4(a) shows 
an excellent agreement with that of ours, and  
the $\epsilon$ and PA distributions of Kent (1987)
in Fig. 4(a) and (b) show reasonable agreements with those of ours.

\vspace{0.5cm}
\centerline{\bf 2) NGC 205}
The $R$ band profiles of $\epsilon$ and PA of NGC 205 are shown 
in Fig. 5(a) and (b) and are listed in Table 4.  
In the region of $R >25$\arcsec, $\epsilon$ increases 
from 0.3 (at $R \sim$ 25\arcsec) to 0.5 (at $R \sim$ 180\arcsec) and 
PA remains at $\sim -15^\circ$, 
while in $R <25 \arcsec$ both $\epsilon$ and PA vary significantly 
($0.00 < \epsilon < 0.52$ and $-41^\circ < PA< 110^\circ$). 
Fig. 5(a) and 5(b) show that there are good agreements 
among Kent (1987)'s, Peletier (1993)'s and our data.

It seems that the innermost stars in $R <25 \arcsec$ 
among the young blue stars in $R <50$\arcsec (Fig. 3(b)) are the populations
that make $\epsilon$ and PA vary significantly 
in the innermost region of $R <25$\arcsec.  
The ellipticity increase from 0.3 to 0.5 at $R >25 \arcsec$ 
might be due to the fact that the stars in the far outer area 
have been affected more by the tidal force of M31 
than those in the inner region(Hodge 1973, Bender \etal 1991).

\vspace{2cm}
\begin{minipage}[c]{8.0cm}
   \epsfxsize=8.28cm
   \epsfysize=15.0cm
   \epsfbox{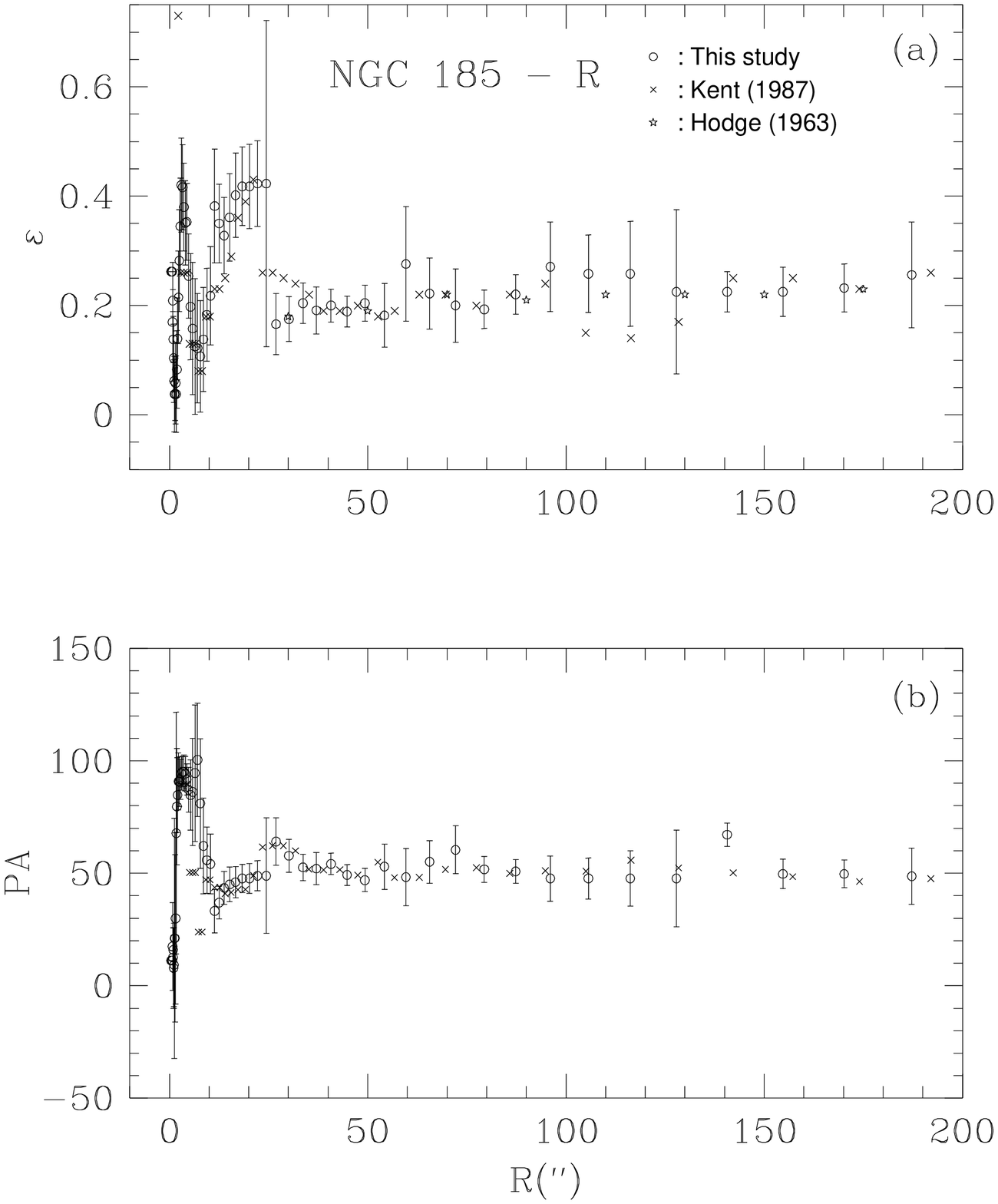}
\end{minipage} \hspace{-0.01cm}
\begin{minipage}[c]{8.0cm}
   \epsfxsize=8.28cm
   \epsfysize=15.0cm        
   \epsfbox{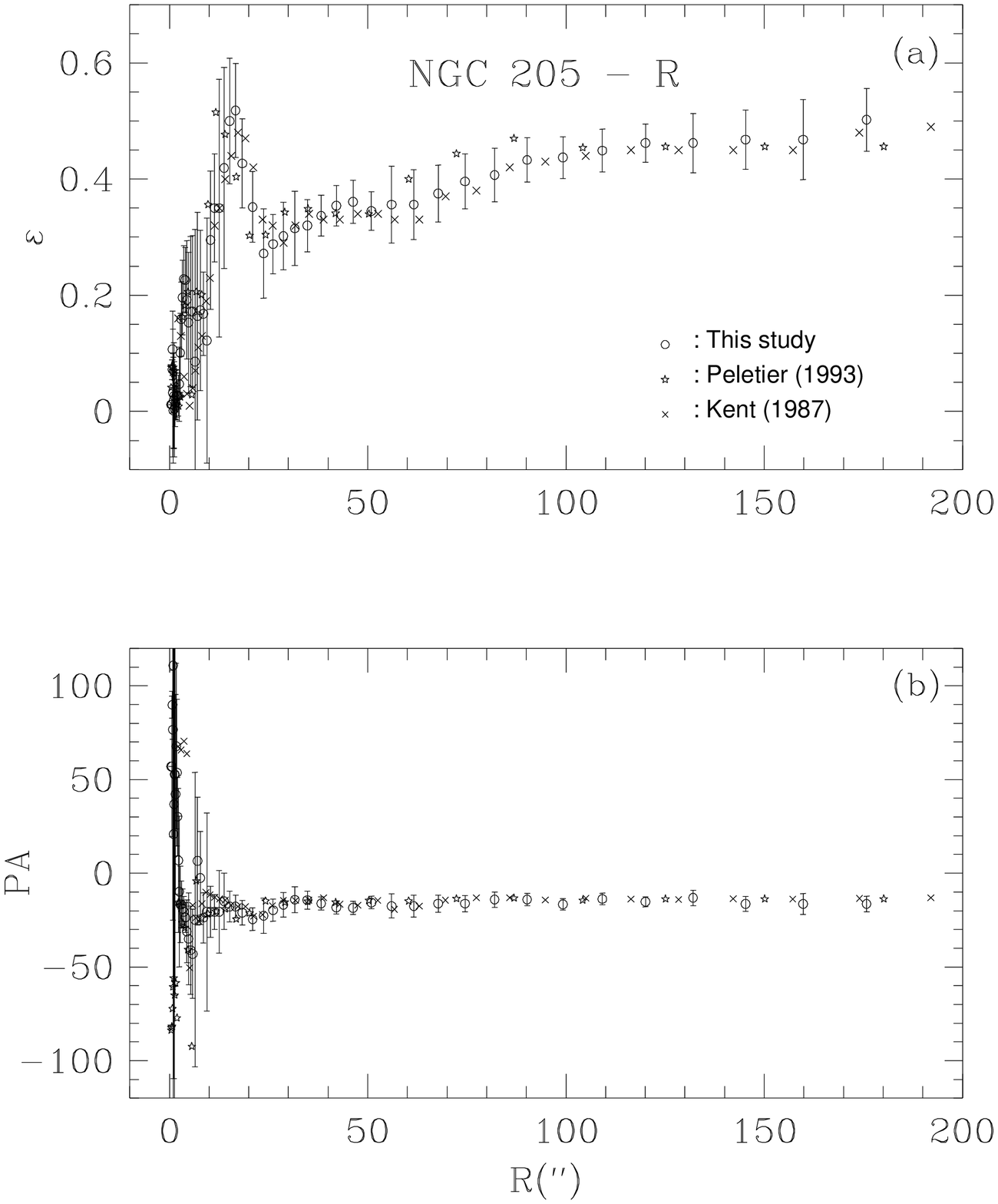}
\end{minipage}
\vspace{2.5cm}

\hspace{0.2cm} {\footnotesize \parbox[t]{8cm}{{\bf Fig. 4.} 
Ellipticity {\bf (a)} and position angle {\bf (b)} profiles of NGC 185 
are shown over the effective radius.  
Ellipticity profiles of Hodge (1963) are shown by $\star$ symbol
and ellipticity and position angle profiles of Kent (1987) 
by $\times$ symbols.}}

\vspace{-1.68cm}
\hspace{9.0cm} {\footnotesize \parbox[t]{8cm}{{\bf Fig. 5.} 
Ellipticity {\bf (a)} and position angle {\bf (b)} profiles 
of NGC 205 are shown over the effective radius.  
Ellipticity and position angle profiles
of Kent (1987) and Peletier (1993) are shown by 
$\times$ and $\star$ symbols, respectively.}}

\vspace{0.5cm}

\section{DISCUSSIONS}

The primary results we obtained in previous section are 
summarized as follows:
(a) the colors of NGC 185 and NGC 205
get bluer inward in the central regions, while they are almost
constant in the outer regions;
(b) the colors of NGC 205 show a reversal of the gradient at $R \approx 
1\arcsec.1$, inside which the colors get redder inward; and
(c) the ellipticity and position angle of the two galaxies
vary significantly in the central region.
These results are believed to be due to the presence of blue stellar
populations in the central regions of the two galaxies
(Baade 1944, 1951; Hodge 1963, 1973; Price \& Grasdalen 1983; Price 1985;
Peletier 1993; Lee \etal 1993; and Lee 1996).
The lack of color gradients and red colors in the outer regions indicate
that few stars were formed recently in the outer regions. 
We estimate approximately the total amount of the blue stellar population
in the central regions of NGC 185 and NGC 205,
and discuss the nuclear color gradient of NGC 205 
and the distribution of dust clouds in NGC 185 and NGC 205
in the following.

{\tiny
\begin{center}
{\normalsize \centerline
   {{\bf Table 4.} Structural parameters of NGC 185 and NGC 205.}}
\begin{tabular}{rccr|rccr}
\tableline
\tableline
NGC 185 &&&& NGC 205 &&&\\
\tableline
$R$(\arcsec)& $\epsilon$  & $\sigma_{\epsilon}$  & PA  &
$R$(\arcsec)& $\epsilon$  & $\sigma_{\epsilon}$  & PA\\
\tableline
0.44&   0.26&   0.00&   11.17&   0.40&   0.01&   0.00&   57.04\\
0.65&   0.26&   0.00&   11.17&   0.60&   0.01&   0.00&   57.04\\
0.72&   0.26&   0.00&   11.17&   0.65&   0.01&   0.01&   57.04\\   
0.79&   0.17&   0.11&   17.45&   0.72&   0.07&   0.02&   89.89\\  
0.87&   0.21&   0.02&   12.51&   0.79&   0.11&   0.07&   89.89\\
0.96&   0.14&   0.04&   15.98&   0.87&   0.03&   0.11&   76.57\\
1.05&   0.10&   0.06&   7.89&   0.96&   0.00&   0.09&   110.83\\   
1.16&   0.06&   0.04&   9.25&   1.05&   0.02&   0.08&   20.83\\
1.27&   0.04&   0.07&   21.06&   1.16&   0.03&   0.09&   36.80\\
1.40&   0.04&   0.05&   21.06&   1.27&   0.00&   0.08&   52.79\\
1.54&   0.06&   0.08&   29.96&   1.40&   0.01&   0.04&   52.79\\
1.70&   0.04&   0.07&   67.84&   1.54&   0.01&   0.03&   42.12\\
1.86&   0.08&   0.07&   79.63&   1.70&   0.03&   0.04&   67.79\\
2.05&   0.14&   0.08&   84.84&   1.86&   0.03&   0.04&   53.69\\   
2.26&   0.22&   0.09&   90.83&   2.05&   0.03&   0.02&   30.19\\
2.48&   0.28&   0.09&   90.83&   2.26&   0.03&   0.04&   6.90\\
2.73&   0.34&   0.09&   91.74&   2.48&   0.05&   0.06&   -9.78\\
3.00&   0.42&   0.09&   94.42&   2.73&   0.10&   0.07&   -16.64\\
3.30&   0.42&   0.08&   95.13&   3.00&   0.16&   0.06&   -16.64\\
3.63&   0.38&   0.08&   95.13&   3.30&   0.20&   0.06&   -18.85\\
3.99&   0.35&   0.08&   94.23&   3.63&   0.23&   0.06&   -20.52\\
4.40&   0.35&   0.07&   91.66&   3.99&   0.23&   0.06&   -23.63\\
4.84&   0.25&   0.08&   87.14&   4.40&   0.19&   0.10&   -31.04\\
5.32&   0.20&   0.10&   84.75&   4.84&   0.15&   0.12&   -34.98\\
5.85&   0.16&   0.12&   86.20&   5.32&   0.17&   0.13&   -41.04\\
6.43&   0.12&   0.12&   94.49&   5.85&   0.17&   0.13&   -42.99\\
7.08&   0.12&   0.10&   100.36&   6.43&   0.09&   0.23&   -24.76\\
7.78&   0.11&   0.10&   81.00&   7.08&   0.16&   0.18&   6.55\\
8.56&   0.14&   0.09&   62.14&   7.78&   0.17&   0.14&   -2.54\\
9.42&   0.18&   0.09&   55.84&   8.56&   0.17&   0.07&   -23.73\\  
10.36&   0.22&   0.09&   54.11&   9.42&   0.12&   0.21&   -20.62\\
11.40&   0.38&   0.10&   33.22&   10.36&   0.29&   0.12&   -20.62\\
12.54&   0.35&   0.07&   36.92&   11.40&   0.35&   0.09&   -20.62\\
13.79&   0.33&   0.07&   43.51&   12.54&   0.35&   0.22&   -20.62\\
15.17&   0.36&   0.08&   45.06&   13.79&   0.42&   0.17&   -14.97\\
16.69&   0.40&   0.08&   46.07&   15.17&   0.50&   0.11&   -17.74\\
18.36&   0.42&   0.07&   47.69&   16.69&   0.52&   0.08&   -17.74\\
20.19&   0.42&   0.08&   47.69&   18.36&   0.43&   0.08&   -21.10\\
22.21&   0.42&   0.08&   48.93&   20.98&   0.35&   0.06&   -24.48\\
24.43&   0.42&   0.30&   48.93&   23.76&   0.27&   0.08&   -22.64\\
26.88&   0.17&   0.06&   64.08&   26.13&   0.29&   0.05&   -19.72\\
30.14&   0.17&   0.04&   57.76&   28.74&   0.30&   0.06&   -16.74\\
33.68&   0.20&   0.04&   52.66&   31.62&   0.31&   0.06&   -14.12\\
37.04&   0.19&   0.04&   52.08&   34.78&   0.32&   0.05&   -14.32\\
40.74&   0.20&   0.03&   54.17&   38.26&   0.34&   0.04&   -16.09\\
44.81&   0.19&   0.03&   49.23&   42.08&   0.35&   0.04&   -18.23\\
49.30&   0.20&   0.03&   46.95&   46.29&   0.36&   0.04&   -18.45\\
54.23&   0.18&   0.06&   52.91&   50.92&   0.34&   0.03&   -15.59\\
59.65&   0.28&   0.10&   48.26&   56.01&   0.36&   0.07&   -17.46\\
65.61&   0.22&   0.06&   55.08&   61.61&   0.36&   0.06&   -17.46\\
72.17&   0.20&   0.07&   60.42&   67.78&   0.38&   0.05&   -16.20\\
79.39&   0.19&   0.04&   51.73&   74.55&   0.40&   0.05&   -16.20\\
87.33&   0.22&   0.04&   50.79&   82.01&   0.41&   0.05&   -14.02\\
96.06&   0.27&   0.08&   47.68&   90.21&   0.43&   0.04&   -14.02\\
105.67&   0.26&   0.07&   47.68&   99.24&   0.44&   0.04&   -16.58\\
116.23&   0.26&   0.10&   47.68&   109.16&   0.45&   0.04&   -13.68\\  
127.86&   0.22&   0.15&   47.68&   120.07&   0.46&   0.03&   -15.17\\
140.64&   0.22&   0.04&   67.13&   132.08&   0.46&   0.05&   -13.09\\
154.70&   0.22&   0.05&   49.73&   145.29&   0.47&   0.05&   -16.35\\
170.17&   0.23&   0.04&   49.73&   159.82&   0.47&   0.07&   -16.35\\
187.19&   0.26&   0.10&   48.69&   175.81&   0.50&   0.05&   -16.35\\
\tableline
\end{tabular}
\end{center}
}

\subsection{Light Excess in the Central Regions}
\centerline{\bf 1) NGC 185}

We have fitted the {\it V}-band surface brightness profiles of NGC 185 
with King (1966) profiles in Fig. 6(a). 
Fig. 6(a) shows that the surface brightness profiles of NGC 185 are well
fitted with two components, rather than one component.
The shaded area in the central region of Fig. 6(a) 
(labeled $'$C$'$, $R <$ 38\arcsec) might be due to
 the blue stars in the 
central area (see Sec. III.2. and Fig. 3(a); Hodge 1963). 
We have estimated approximately the light excess of the central region 
to be $\simeq 10^5$ L\sun. 
\vspace{0.6cm}

\centerline{\bf 2) NGC 205}
Fig. 6(b) shows that the surface brightness
profiles of NGC 205 are well fitted with three components:
nuclear region, inner region, and outer region. 
We have calculated the light excesses in the nuclear region ($R <$ 2\arcsec.7) 
to be $2 \times 10^5$ L\sun, and in the inner region 
($R <$ 46\arcsec) to be $\approx 10^5$ L\sun.
Thus, the total light excess resulted from both N and C regions in NGC 205 is
$\simeq 3 \times 10^5$ L\sun.  

\hspace{2cm}
\begin{minipage}[t]{12cm}
   \epsfxsize=12cm
   \epsfysize=5.0cm
   \epsfbox{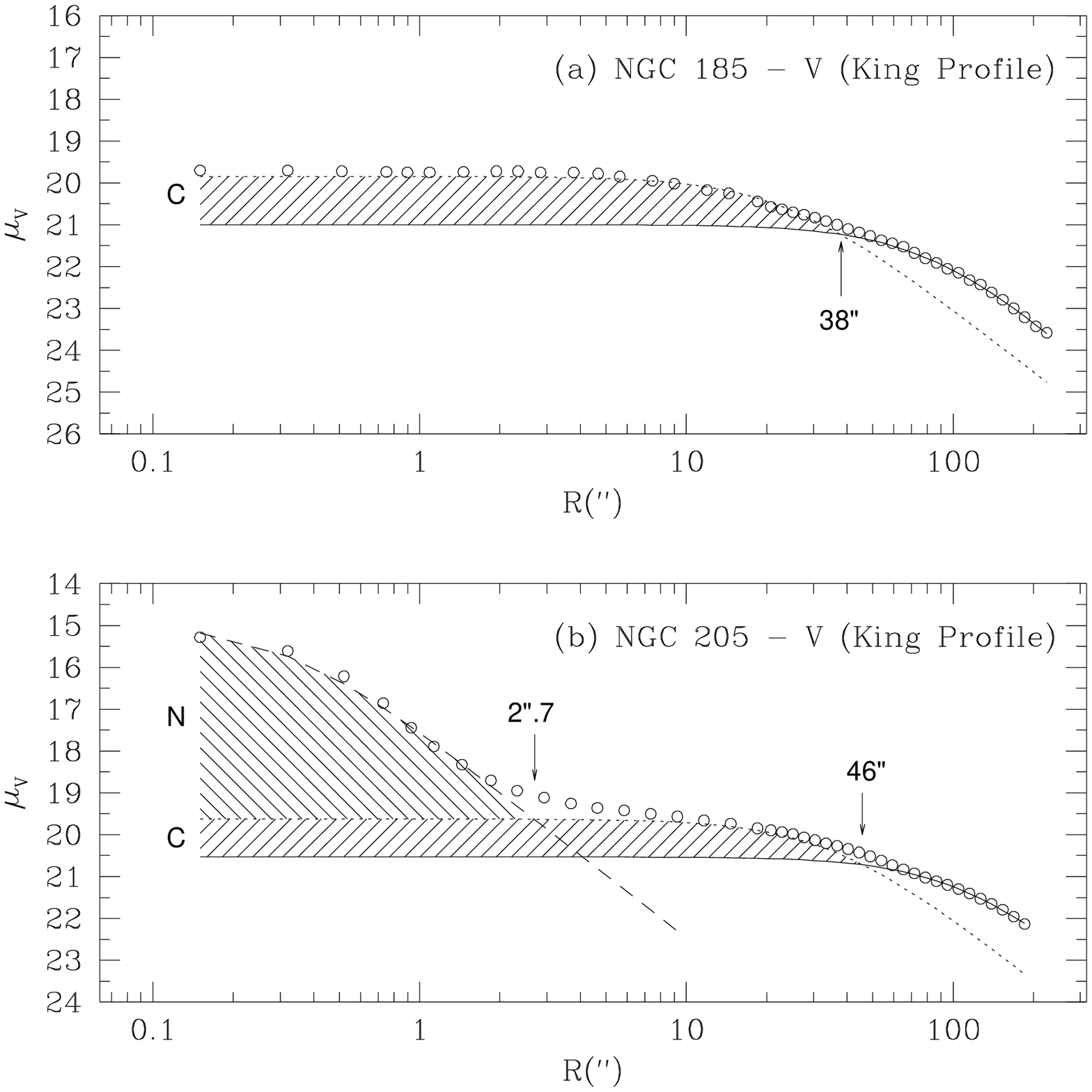}
\end{minipage} \hfill

\hspace{-0.8cm} {\footnotesize \parbox[t]{17.5cm}{{\bf Fig. 6.} 
King model fitting for the {\it V} surface brightness profiles 
of {\bf (a)} NGC 185 and {\bf (b)} NGC 205.
NGC 185 is fitted by two components and the region of the
central light excess ($R <$ 38$\arcsec$) is shaded and labeled C.  
NGC 205 is fitted by three components and the two regions
of light excesses are shaded and labeled as N (nuclear part, $R <$ 2\arcsec.7)
and C (central part, $R <$ 46\arcsec).}}

\subsection{Color Gradient in the Nuclear Region of NGC 205$^1$}

\footnotetext[1]{Based on observations made with the NASA/ESA Hubble Space Telescope,
obtained from the data archive at the Space Telescope Science Institute.  STScI is 
operated by the Association of Universities for Research in Astronomy, Inc. under
NASA contract NAS 5-26555.}

The color profiles of NGC 205 in Fig. 3(b) show that all the colors get redder
in the nuclear region of this galaxy.
To investigate this color inversion of the nucleus color in detail, 
we have analyzed the images of the central region in the $HST$ archive data.

We have performed surface photometry on $HST$ FOC f/48 + F150W + F130LP
images used by Bertola \etal (1995) and $HST$ WFPC2 F555W images used by Jones \etal (1996). 
Jones \etal also obtained far-ultraviolet (F160BW) images of NGC 205, but the
quality of the images is not good enough to derive surface photometry from them,
so that we did not use them in this study.

The resulting surface brightness profiles ($\mu_{1500}$ and $\mu_{5550}$)
and color profiles ($\mu_{1500} - \mu_{5550}$) are shown in Fig. 7(a) and (b),
respectively. A comparison of the observed {\it V}-band surface brightness profiles in our study
and Jones \etal (1996) shows that both agree very well.

We have tried to calibrate the far-ultraviolet photometry following 
the description given by Bertola \etal (1995). However, we have obtained 
as the color of the central region within the circular aperture of the radius of 
0\arcsec.273 , $m(1500)-m(5550)$ = 0.67 mag, 
which is much bluer than the value given by Jones \etal (1996), 1.20. 
The reason for this difference is not known at the moment.
So we have used the far-ultraviolet photometry 
only in the relative sense in this study.

In Fig. 7(a) we also show the surface brightness profiles which are obtained from the
deconvolved images created using Lucy-Richardson deconvloution in IRAF/STSDAS.
The $\mu_{5550}$ profile based on the deconvolved image in this study is slightly 
brighter in the central region than that of Jones \etal (1996). 
This difference appears to be due to the fact that
we used a bright star in the image for deriving a point spread function, while
Jones \etal (1996) used a library point spread function given by Krist \& Burrows (1995).

In Fig. 7(b) we have plotted the color profiles based on the original images
and deconvolved images. 
Fig. 7(b) shows that the color gets bluer inward until about 1\arcsec, inside which
the color gets redder rapidly inward until about 0\arcsec.2. 
The color stays almost constant inside of $R = 0\arcsec.2$
at the value $\sim1.6$ mag redder than the color at $R \approx 1\arcsec$
on original images.
This radius corresponds to the core radius
of the nucleus estimated on the $HST$ image (Jones \etal 1996).
These results are consistent with those based on ground-based data.

\vspace{0.5cm}
\hspace{3cm}
\begin{minipage}[t]{14cm}
   \epsfxsize=10cm
   \epsfysize=3.0cm
   \epsfbox{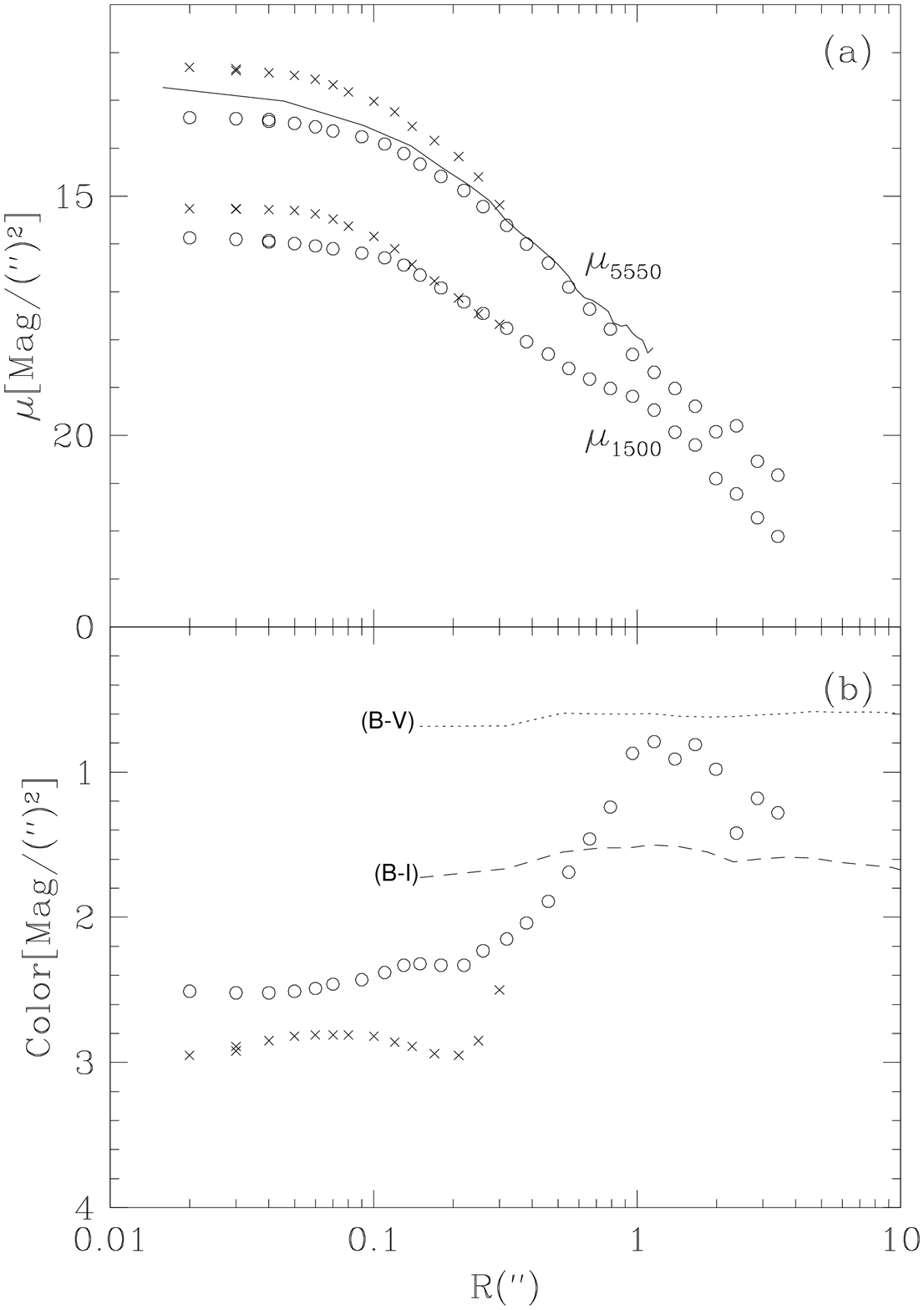}
\end{minipage} \hfill

\hspace{-0.8cm} {\footnotesize \parbox[t]{17.5cm}{{\bf Fig. 7.}
{\bf (a)} Surface brightness profiles of $\mu_{1500}$ 
and $\mu_{5550}$ from $HST$ images for the nuclear region of NGC 205.
Surface brightnesses from images before deconvolution are shown in circles,
and those from images after deconvolution are in crosses.
Surface brightnesses of Jones \etal (1996) obtained 
from the images deconvolved by themselves are plotted as a solid line.
{\bf (b)} Color profiles ($\mu_{1500} - \mu_{5550}$) 
for the nuclear region of NGC 205.
Colors from the original images are shown in circles,
and those from the deconvolved images are in crosses.
$B-V$(dotted line) and $B-I$(dashed line) colors 
are also shown for comparison.}}

Thus these results show clearly that the core region of the nucleus of NGC 205 is 
distinctively redder than the outer part of the nucleus.
This result may be interpreted either that there is a significant reddening at the center or
that there is a component at the center distinguishable from the neighbor region of
the nucleus.
The smooth surface brightness profiles in the central region rule out 
the first possibility. 
Then there remains the second interpretation as a good possibility.
The trend of being bluer inward in the central region indicates that there have been
star formations in the central region of NGC 205 until recently as discussed in Lee (1996).
However, the core part of the nucleus is probably not related with the recent star
formations. 

Previous spectroscopic observations of the nucleus of NGC 205 showed 
the presence of Balmer lines in the optical spectra 
and a rather flat continuum(or some untraviolet upturn)
in the ultraviolet spectra obtained using IUE, which have been considered 
as indications of recent star formation
(Da Costa \& Mould 1988, Burstein \etal 1988, Wilcots \etal 1990).
However, these spectra were obtained using large apertures of 5\arcsec to $25\arcsec$
and these phenomena are probably due to the central region except for the very central region
of NGC 205. Spectroscopy of the central 0\arcsec.4 is needed to confirm our
reasoning.   

\subsection{Distribution of Dust Clouds}

We have investigated the distribution of dust clouds using the images
which are created by subtracting a model galaxy based on the surface photometry
from the original images.

\vspace{0.8cm}
\centerline{\bf 1) NGC 185}
Fig. 8(a) shows the image which was made by subtracting a model galaxy 
from the original CFHT $B$ image of NGC 185. The model galaxy was created
using the ELLIPSE task in IRAF/STSDAS.
Several prominent dust clouds are clearly seen in Fig. 8(a), most of which are
concentrated in the central region of the galaxy. 
Fig. 8(b) shows a schematic diagram of these dust clouds (DCs).

  The most conspicuous cloud is the one at NW direction 
from the center (DC I of Hodge 1963 and Gallagher \& Hunter 1981).  
The second most conspicuous cloud, DC II, is at SE direction of the center, 
starting from the bottom of DC I. DC I and II were previously known
by Hodge (1963). 
There are seen four more dust clouds in the central region of NGC 185: 
DCs III, IV, V, and VI.
Two small and dense clouds, which we named DC IV and DC V, 
are located at NE direction of the center of NGC 185.
Two diffuse and large clouds, which we named DC III and DC VI,
are located at SW direction of DC II and NE direction of DC IV, respectively.
Interestingly the clouds DC I,II and VI are arranged in a bow-shape.
The position of DC II coincides with the position of the peak HI column density,
where CO molecular gas was also detected (Young \& Lo 1997).

\vspace{1cm}

\centerline{\bf 2) NGC 205}
  In Fig. 9(a), we have shown the image of NGC 205 made 
by subtracting the model image from the original {\it B} image.
In Fig. 9(b), we have shown a schematic diagram of dust clouds and bright stars
which are in Fig. 9(a).  
Besides the twelve dust clouds reported by Hodge (1973), 
which are plotted as solid lines in Fig. 9(b), 
three more diffuse absorption regions, 13, 1A, and 3A 
are seen in Fig. 9(a) and (b).  
DC 13 appears to be denser than DC 2 and to be connected to the nucleus of NGC 205.
Two diffuse dust clouds, DC 1A and DC 3A,
are seen around DC 1 and DC 3, respectively.  
Hodge (1973) delineated only the most eminent parts of DC 1A and DC 3A as DC 1 and DC 3,
respectively, in his Fig. 22.

  We have plotted HI column density map of Young \& Lo (1997) by contours in Fig. 9(a). 
Fig. 9(a) shows that the dust clouds and HI have almost the same distributions 
in shape and extent.
The places where the prominent dust clouds (DCs 11, 1A, and 12) exist are
where HI distribution shows local peaks and makes a ridge connecting the peaks.

\vspace{1cm}
\begin{minipage}[t]{12.0cm}
   \epsfxsize=8.28cm
   \epsfysize=12.0cm
   \epsfbox{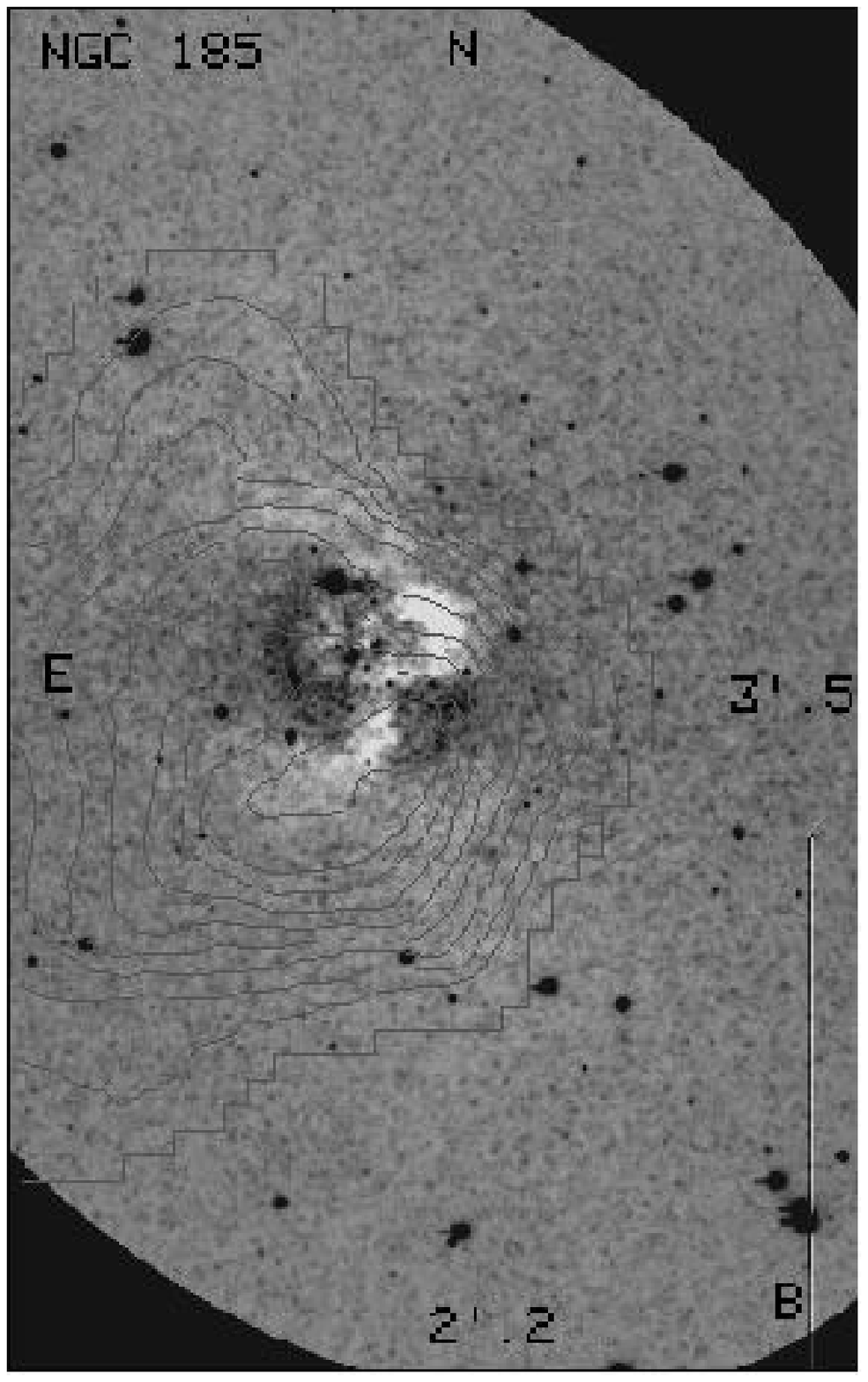}
\end{minipage} \hfill
\begin{minipage}[t]{12.0cm}
   \epsfxsize=8.28cm
   \epsfysize=12.0cm
   \epsfbox{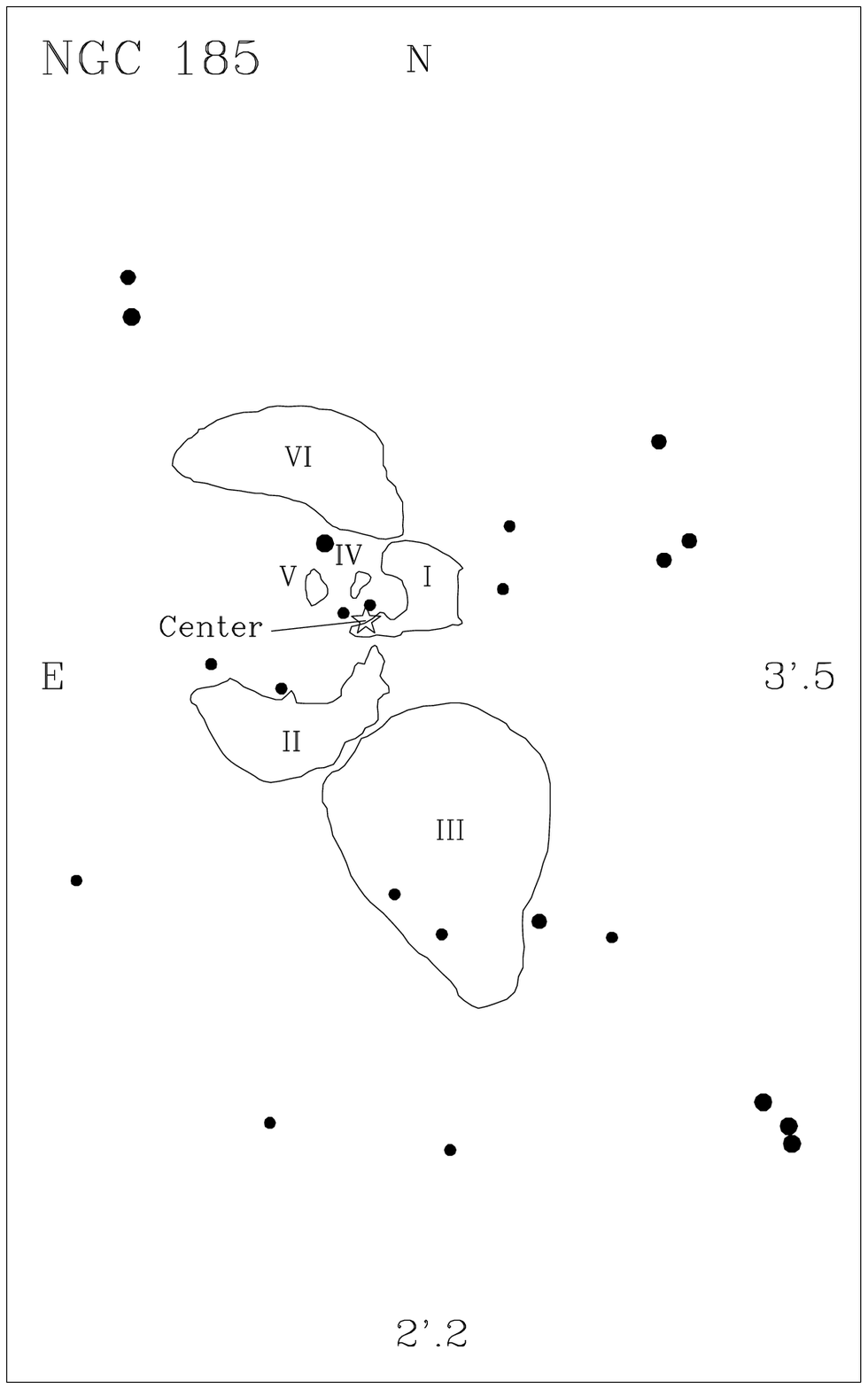}
\end{minipage}

(a) \hspace{8.0cm} (b)

\hspace{-0.8cm} {\footnotesize \parbox[t]{17.5cm}{{\bf Fig. 8. (a)} 
A greyscale map of the CFHT $B$ CCD image of NGC 185 used in Lee \etal (1993).  
North is at the top and east is to the left.  
The size of the field is $2\arcmin$.2 $\times 3\arcmin$.5.
The model image of NGC 185  
was subtracted from the original image.  
Several prominent dust clouds are seen clearly.  
HI column density map of Young \& Lo (1997) is plotted by the contours.
{\bf (b)} A schematic diagram showing noticeable dust clouds
and bright stars in Fig. 8(a) for NGC 185.  
The center of NGC 185 is shown by a $\star$ symbol.
}}

\section{SUMMARY AND CONCLUSIONS}
  We have presented surface photometry of 
two peculiar dwarf elliptical galaxies NGC 185 and NGC 205.  
The primary results are summarized below.

\begin{minipage}[t]{12.0cm}
   \epsfxsize=8.28cm
   \epsfysize=12.0cm
   \epsfbox{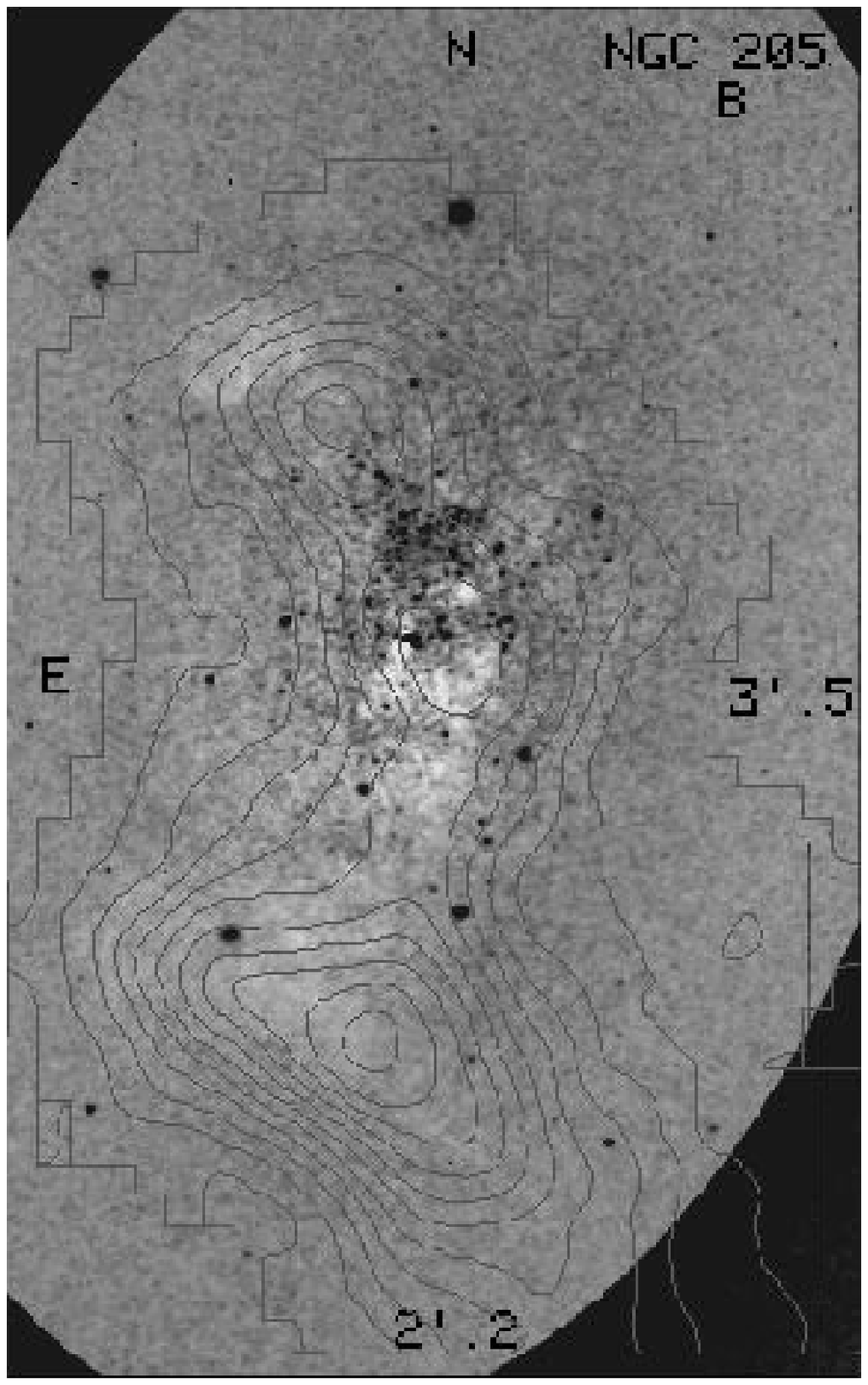}
\end{minipage} \hfill
\begin{minipage}[t]{12.0cm}
   \epsfxsize=8.28cm
   \epsfysize=12.0cm
   \epsfbox{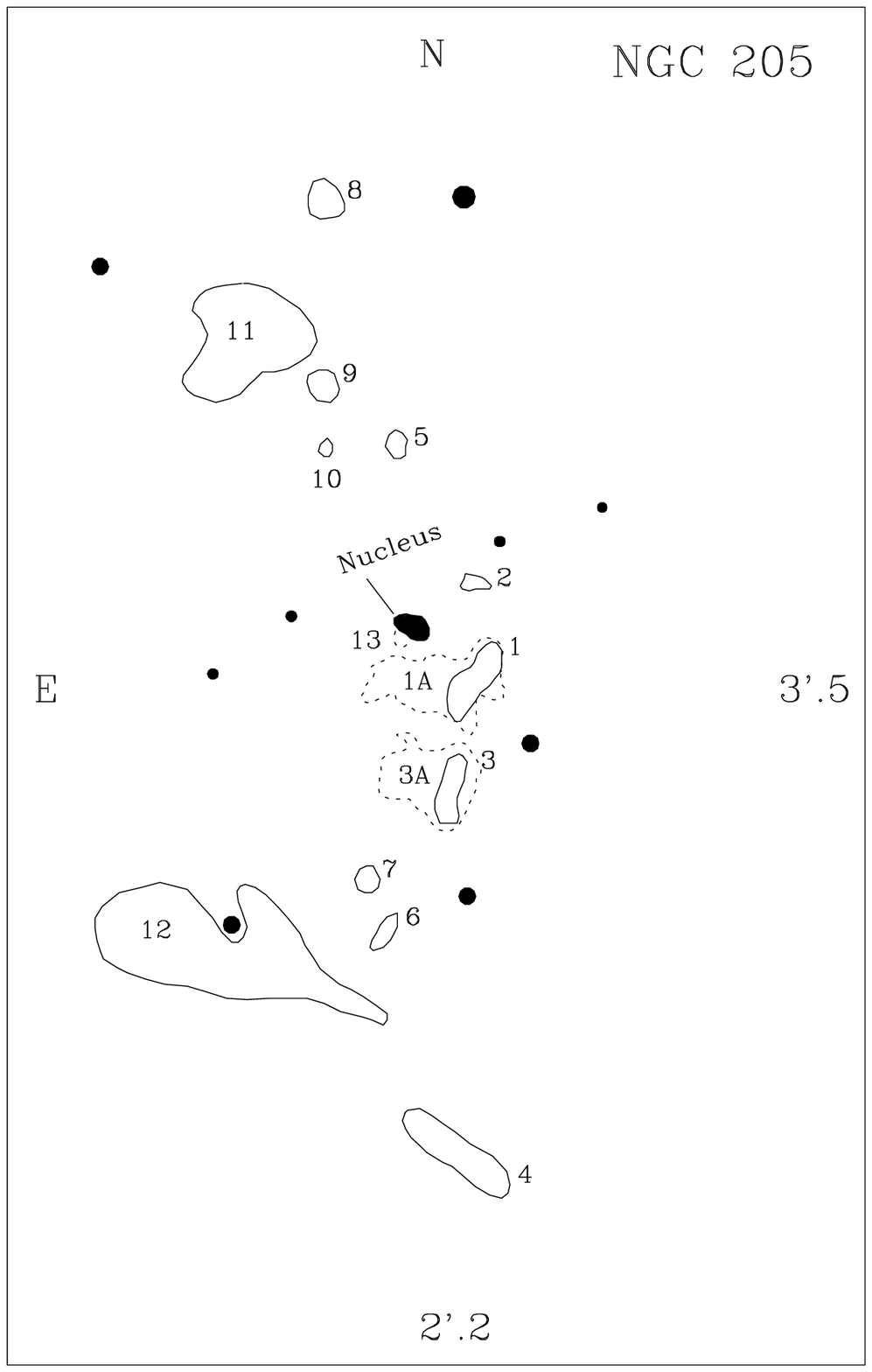}
\end{minipage}

(a) \hspace{7.9cm} (b)

\hspace{-0.8cm} {\footnotesize \parbox[t]{17.5cm}{{\bf Fig. 9. (a)} 
The greyscale map of the $B$ CCD image of NGC 205 obtained
using the CFHT 3.5m telescope.  North is at the top and east is to the
left.  The size of the field is $2\arcmin$.2 $\times 3\arcmin$.5.
The model image of NGC 205  
was subtracted from the original image.  Some of prominent dust clouds
are seen easily in this figure.  HI column density map of Young \& Lo (1997)
is plotted by the contours. 
{\bf (b)} A schematic diagram showing some of the dust clouds.
Some bright stars and globular clusters are shown.
Three additional dust clouds at the southern part of the nucleus are 
labeled as 13, 1A, and 3A.}} 
\vspace{1cm}

1. The surface brightness profiles of NGC 185 are
flat in the central region and decrease smoothly outward,
while those of NGC 205 brighten rapidly in the nuclear region,
become flat in the central region and
decrease smoothly outward in the outer region.

2. The colors of NGC 185 get bluer inward in $R < 25 \arcsec$ 
and remain almost flat at $R > 25 \arcsec$.  
The colors of NGC 205 also get bluer inward at $1\arcsec < R < 50\arcsec$ 
and remain nearly flat outside of $R \sim$ 50\arcsec,
while the nuclear ($R < 1\arcsec$) colors get redder inward.

3. The radial profiles of ellipticity and the position angle of NGC 185 remain flat 
($\sim$0.2 and $\sim 50^\circ$, respectively) outside of $R \sim$ 25\arcsec, 
while both show significant variations within that radius.  
The ellipticity of NGC 205 increases 
from 0.3 at $R \sim$ 25$\arcsec$ to 0.5 at $R \sim$ 180$\arcsec$ and 
the position angle remain flat ($\sim -15^\circ$) in that range, 
while both the ellipticity and the position angle vary wildly in $R <$ 25\arcsec.

4. The excesses of light are estimated to be $\simeq 10^5$ L\sunn for NGC 185, 
and $\simeq 3 \times 10^5$ L\sunn for NGC 205.  
These light excesses seem to be due to the young stars 
in the central regions of the two galaxies.

5. Several new dust clouds are found in NGC 185 and NGC 205.

\acknowledgments
D. Heath Jones is thanked for providing us with his light profile data of NGC 205,
and Lisa M. Young is thanked for sending us HI maps of NGC 185 and NGC 205.
This research is supported in part by the Korean Ministry of Education through Research
Fund BSRI-97-5411.

\end{document}